# A Numerical Treatment of the Rf SQUID: II. Noise Temperature

**Reinhold Kleiner[1], Dieter Koelle[1] and John Clarke[2]**

[1]Physikalisches Institut-Experimentalphysik II, Universität Tübingen, 72076 Tübingen, Germany

[2]Department of Physics, University of California, Berkeley, California 94720-7300,

and

Materials Sciences Division, Lawrence Berkeley National Laboratory, Berkeley, California 94720

E-mail: kleiner@uni-tuebingen.de

*We investigate rf SQUIDs (Superconducting QUantum Interference Devices), coupled to a resonant input circuit, a readout tank circuit and a preamplifier, by numerically solving the corresponding Langevin equations and optimizing model parameters with respect to noise temperature. We also give approximate analytic solutions for the noise temperature, which we reduce to parameters of the SQUID and the tank circuit in the absence of the input circuit. The analytic solutions agree with numerical simulations of the full circuit to within 10%, and are similar to expressions used to calculate the noise temperature of dc SQUIDs. The best device performance is obtained when $\beta'_L \equiv 2\pi L I_0 / \Phi_0$ is 0.6 - 0.8; L is the SQUID inductance, $I_0$ the junction critical current and $\Phi_0$ the flux quantum. For a tuned input circuit we find an optimal noise temperature $T_{N,opt} \approx 3Tf / f_c$, where T, f and $f_c$ denote temperature, signal frequency and junction characteristic frequency, respectively. This value is only a factor of 2 larger than the optimal noise temperatures obtained by approximate analytic theories carried out previously in the limit $\beta'_L << 1$. We study the dependence of the noise*

*temperature on various model parameters, and give examples using realistic device parameters of*

*the extent to which the intrinsic noise temperature can be realized experimentally.*

## 1. INTRODUCTION

As a result of the continuing progress in computing speed, numerical simulation and noise optimization, a nonlinear electronic circuit such as the rf SQUID (Superconducting Quantum Interference Device) coupled to a tank circuit has become an accessible problem. Nonetheless, the optimization of all parameters may still be time-consuming. The noise performance of rf SQUIDs has been investigated analytically in the limit of either very large or very small values of the inductance parameter $\beta'_L \equiv 2\pi I_0 L / \Phi_0$ (where $I_0$ is the junction critical current, $L$ is the inductance of the SQUID loop and $\Phi_0$ is the flux quantum), assuming that the inductive coupling between the SQUID loop and the tank circuit is weak.[1] The SQUID is coupled to a readout tank circuit, consisting of an inductance $L_T$, capacitance $C_T$ and resistance $R_T$ in series, via a mutual inductance $M = \alpha (L L_T)^{1/2}$ (Fig. 1). It turns out, however, that the optimal device parameters are outside the validity of these approaches. In a previous publication [2] (which we suggest the reader study before tackling this paper) we calculated and optimized the noise energy $\varepsilon$ of rf SQUIDs by numerically solving the corresponding Langevin equations. We found that the best values were obtained for $\beta'_L \approx 1$. For values of the noise parameter $\Gamma \equiv 2\pi k_B T / I_0 \Phi_0$ below about 2, we found the normalized noise energy $e = \varepsilon /(2 k_B T \Phi_0 / I_0 R)$ ($R$ is the junction resistance) to be $0.5 \exp(\Gamma \beta'_L)$, albeit for values of $\alpha$ well above 0.2. Perhaps surprisingly the best values of $e$ were only a factor of 2 above the prediction for the dispersive rf SQUID ($\beta'_L < 1$) obtained by extrapolating the analytic results obtained, for example, by Danilov, Likharev and Snigirev,[3] for $\beta'_L \ll 1$ to values of $\beta'_L$ near unity. The numerical simulations in Ref. [2], however, neglected both the back-action of the rf SQUID on an input circuit and the back-action of the preamplifier following the tank circuit. A complete theory should include both the input circuit and the preamplifier. In this case the noise temperature $T_N$ is a more suitable figure of merit than the noise energy. One defines $T_N$ by considering the Nyquist noise of a resistor at temperature $T_i$ in the input circuit (Fig. 1). In the classical regime, when $T_i$ is increased from zero to its value at which the output noise power doubles, that value is equal to $T_N$. For a tuned input circuit, $T_N$ becomes independent of its parameters once they are optimized to minimize[3] $T_N$. During the 1970s, the noise temperature of rf SQUID circuits was considered by several authors in both the dissipative[3, 4, 5] and dispersive regimes[3] $\beta'_L \gg 1$ and $\beta'_L \ll 1$, respectively. The most complete description, to our knowledge, is that of Danilov, *et al.*[3] For $\Gamma \ll 1$, these authors studied the noise performance of both dc and rf SQUIDs in the limits $\beta'_L \gg 1$ and $\beta'_L \ll 1$. In the dispersive regime, they found the minimum noise temperature of the rf SQUID coupled to a tuned input circuit was given by

$$T_{N,min}/T \approx 2.5(1+T_{tank}/T\alpha^2 Q\beta_L' f_d)f/f_c \quad . \tag{1}$$

Here, $f_c = I_0 R / \Phi_0$ is the characteristic frequency of the junction, $f$ is the measurement frequency, $f_d$ is the rf drive frequency, $Q$ denotes the tank circuit quality factor and $T_{tank}$ is the effective temperature of the tank circuit resistor, including the noise contribution of the preamplifier. The approximate factor 2.5 results from $2^{1/2}/J_1(1.8)$ , where $J_1$ is the first-order Bessel function. When the second term in parentheses can be neglected one obtains $T_{N,min}/T \approx 2.5f/f_c$, which for, say, $T = 4.2$ K and $f_c = 100$ GHz, yields $T_{N,min}/f \approx 100 \,\mu$K /MHz. We shall see below that the best predicted intrinsic noise temperatures are only slightly above this value, and are achieved for $\beta'_L$ of 0.6-0.8 and $\alpha$ above about 0.2. Furthermore, the noise temperature remains low up to $\Gamma = 1$ or even higher. Our simulations are mostly for $Q = 100$, which is sufficient to achieve low values of $T_N$. As for the case of the noise energy discussed in Ref. 2, we were not able to duplicate the inverse scaling with $\alpha^2 Q f_d$, indicated in Eq. (1).

The paper is organized as follows. In Sec. 2, we introduce the full rf SQUID circuit and the Langevin equations describing it. While we solve this set of equations only for some cases, we minimize the noise temperature analytically for a tuned input circuit optimized for the parameters of this circuit, $T_{N,opt} = \pi(\widetilde{S}_{VT}\widetilde{S}_J - \widetilde{S}_{VJ}^2)^{1/2} f / V_\Phi$. Here, $\widetilde{S}_{VT}$ (f) and $\widetilde{S}_J$ (f) denote the spectral density of the low frequency voltage $V_T$ across the tank circuit and the circulating current $J$ in the SQUID loop in the absence of an input circuit, $\widetilde{S}_{VJ}$ (f) is the cross correlation between the Fourier transforms of these quantities and $V_\Phi$ is the modulus of the SQUID flux-to-voltage transfer function $(\partial V_T / \partial \Phi_{ext})_{I_T}$; $\Phi_{ext}$ is the flux applied to the SQUID and $I_T$ is the rf drive current of the tank circuit. The expression for $T_{N,opt}$ follows from a low frequency analysis of both dc[3,6] and rf SQUID circuits,[3] and is re-derived in Appendix A1. In Sec. 3, we optimize $T_{N,opt}$ with respect to various model parameters, assuming a narrowband readout scheme following the tank circuit. Section 4 is devoted to broadband readout schemes and the dependence of $T_{N,opt}$ on the drive frequency. Section 5 contains our conclusions. In Appendix A2, the analytic results derived in Appendix A1 are compared to numerical simulations. Appendix A3 contains a list of symbols.

## 2. MODEL

The circuit, shown in Fig.1, consists of three units coupled inductively: the input circuit, the SQUID loop and the tank circuit. Before we describe these components, however, it is convenient first to discuss the preamplifier connected across the tank circuit.

We assume that the preamplifier is based on a high electron mobility transistor[7,8] (HEMT) and cooled to 4.2 K. We characterize the preamplifier by an input impedance[9] $R_A$ (which we assume to be real), and input voltage and current noise sources $U_{NA}$ and $I_{NA}$ with spectral densities $S_{UA}$ and $S_{IA}$, respectively, at or near the resonant frequency of the tank circuit. For a typical cooled HEMT amplifier with an input impedance matched to 50 $\Omega$, representative noise values are $S_{UA}^{1/2} = 1 \times 10^{-10}$ V Hz$^{-1/2}$ and $S_{IA}^{1/2} = 2 \times 10^{-12}$ A Hz$^{-1/2}$. We assume that the two noise sources are uncorrelated and have white spectral densities. The current noise feeds into the tank circuit and represents a backaction, while the voltage noise source appears only at the output of the preamplifier, amplified by the preamplifier gain. For convenience in our formulation, we parameterise the current noise by the spectral density $S_{IA} = 4k_B T_A / R_A$. We emphasize, however, that the temperature $T_A$ does *not* represent any physical temperature. For the optimal source resistance $R_{opt} = S_{UA}^{1/2} / S_{IA}^{1/2}$, one can easily relate $T_A$ to the optimal noise temperature[10] of the preamplifier, $T_{NA}$. Since $T_{NA} = S_{IA} R_{opt} / 2k_B$, we see that $T_{NA} = 2(R_{opt}/R_A)T_A$. For the noise values listed above, we find $R_{opt} = 50\ \Omega$ and $T_A = 7$ K.

The SQUID is characterized by the dimensionless parameters $\beta'_L$ and $\beta_c = 2\pi I_0 R^2 C / \Phi_0$; $C$ is the junction capacitance. At temperature $T$, $R$ produces a Nyquist current noise $I_N$ with a spectral density $4k_B T / R$, or $4\Gamma = 4 \cdot 2\pi k_B T / I_0 \Phi_0$ in dimensionless units.

The tank circuit is driven by an oscillating current drive $I_T$ with amplitude $I_d$ ($i_d$ in units of $I_0$) at frequency $f_d$ (in units of $f_c$) and consists of an inductor $L_T$, a capacitor $C_T$ and a resistor, $R_T$ representing losses in the tank circuit; $R_A$ appears in parallel with $R_T$. For $R_A \rightarrow \infty$ (as used in Ref. 2 to calculate the noise energy) the tank circuit is characterized by the unloaded quality factor $Q_0 = (L_T / C_T)^{1/2} / R_T$ and the normalized resonance frequency $f_0 = 1/f_c\, 2\pi(L_T C_T)^{1/2}$. For finite values of $R_A$ we define $Q_A^2 = R_A^2 C_T / L_T$ and an effective quality factor $Q_{eff,0} = Q_0 Q_A / (Q_0 + Q_A)$. (This definition of $Q_{eff,0}$ is chosen so that, at resonance (frequency $\omega_r$) and in the absence of the SQUID loop and the input circuit, the linear tank circuit at zero temperature delivers a voltage $U = Z_{real} I = Q_{eff,0} L_T \omega_{res} I$, or, in dimensionless units, $u = \beta'_L f_0 Q_{eff,0} i \gamma_L^2$.)

We assume that $R_T$ produces Nyquist noise at temperature $T_T = T$, corresponding to a voltage source $U_{N,T}$ with spectral density $4k_B T R_T$ [or, in normalized units, $4\Gamma_T \gamma_R = 4 \cdot 2\pi k_B T \gamma_R / I_0 \Phi_0$, where $\gamma_R = R_T / R = \beta'_L f_0 \gamma_L^2 / Q_0$ and $\gamma_L = (L_T / L)^{1/2}$]. At this point, it is convenient to introduce the dimensionless spectral density for the preamplifier current noise $s_{iA} = 4\Gamma_A R / R_A$, where $\Gamma_A \equiv$ $2\pi k_B T_A / I_0 \Phi_0$. Using the result $R_A / R = \beta_L' f_o \gamma_L^2 Q_A$, we can also write the spectral density as $s_{iA} = 4\Gamma_A / Q_A \beta_L' f_o \gamma_L^2$.

At resonance, the tank circuit impedance is given by $Z_r = [R_A^{-1} + (Q_0^2 R_T)^{-1}]^{-1}$. Defining $R_{opt} / Z_r = \kappa$ we obtain the expressions $R_{opt} = \kappa Q_{eff,0} \sqrt{L_T / C_T}$ and $\Gamma_{NA} = 2\kappa \Gamma_A Q_{eff\,0} / Q_A$, with $\Gamma_{NA} = 2\pi k_B T_{NA} / I_0 \Phi_0$. For the normalized spectral densities of the preamplifier voltage and current noise we obtain $s_{uA} = 4\kappa^2 Q_{eff,0}^2 \beta_L' \gamma_L^2 f_0 \Gamma_A / Q_A = 2\kappa Q_{eff,0} \beta_L' \gamma_L^2 f_0 \Gamma_{NA}$ and $s_{iA} = 4\Gamma_A / Q_A \beta_L' \gamma_L^2 f_0 = 2\Gamma_{NA} / \kappa Q_{eff,0} \beta_L' \gamma_L^2 f_0$, respectively. At a fixed value of $\Gamma_{NA}$, $s_{uA}$ and $s_{iA}$ scale as $\kappa$ and $1/\kappa$, respectively, so that $\kappa$ has an optimal value. We performed numerical simulations to optimize the system noise temperature, including $s_{iA}$ and $s_{vA}$, done for $\Gamma_{NA} = 0.05$, $\Gamma = 0.025$, $Q_{eff,0} = 100$, $\alpha = 0.2$ and several values of $f_0$ between 0.5 and 0.01, while optimizing $f_d$, $i_d$, and $\beta_L'$. These simulations[11] yielded only a weak dependence on $\kappa$ in the range 0.1 - 1. Below we consider this range of $\kappa$ as "typical", although we have not performed a systematic investigation varying many model parameters. Below, unless stated otherwise, we ignore the contribution of $s_{UA}$ which simply adds to the voltage output of the tank circuit.

For the case in which the preamplifier weakly damps the tank circuit, for example, $Q_0 = 101$ and $Q_A = 10100$, corresponding to $Q_A / Q_0 = 100$ and $Q_{eff,0} = 100$, we find $T_A \approx T_{NA} Q_A / 2\kappa Q_0$. In this limit, $T_A$ and thus $\Gamma_A$ can take large values: for example, for $\Gamma_A = 10\Gamma = 0.25$ and T = 4.2 K, we find $T_A = 42$ K and $T_{NA} / \kappa = 0.2$ K. Since for $\kappa \lesssim 10$ the corresponding values of $T_{NA}$ are well below the noise temperature of any available semiconductor preamplifier, we use these parameters to represent a "quiet" preamplifier that has essentially no impact on the SQUID readout. A second interesting limit is the matched case $Q_A = Q_0$ for which $\kappa T_A = T_{NA}$. The same set of values for $\Gamma$, $\Gamma_A$ and $T$ leads to $T_{NA} / \kappa = 42$ K, which for $\kappa > 0.3$ or so is well above the noise temperature of good available semiconductor preamplifiers. We refer to this situation as a "noisy" preamplifier that dominates the overall system noise temperature.

The input circuit consists of an inductor $L_i$, a capacitor $C_i$ and a resistor $R_i$ that produces Nyquist noise at temperature $T_i$ (Fig. 1). The (unloaded) quality factor is $Q_{0i} = (L_i / C_i)^{1/2} / R_i$ and the normalized resonance frequency is $f_{0i} = \Phi_0 / 2\pi I_0 R \sqrt{L_i C_i}$. The inductance $L_i$ is coupled to $L$ via the coupling coefficient $\alpha_i$ and to $L_T$ via $\alpha_{iT}$. Later on we further assume that the input circuit resonates at a frequency well below the characteristic frequencies of the SQUID loop and the tank circuit.

We next introduce the equations describing the full circuit. The fluxes through $L_T$, $L_i$ and $L$ are given by $\Phi_T = L_T I_1 + MJ + M_{iT} I_i$, $\Phi_i = L_i I_i + M_i J + M_{iT} I_1$ and $\Phi = \Phi_{ext} + LJ + MI_1 + M_i I_i$, respectively. Here, $M_i = \alpha_i (LL_i)^{1/2}$ is the mutual inductance between the input loop and the SQUID, $M_{iT} = \alpha_{iT} (L_i L_T)^{1/2}$ is the mutual inductance between the input circuit and the tank circuit, and $I_i$ is the current in the input circuit. In dimensionless units, with $\gamma_{Li} = \sqrt{L_i / L}$, $i_I = I_I / I_0$, $j = J / I_0$ and $i_i = I_i / I_0$, we find

$$\varphi_T = \beta'_L \gamma_L [\gamma_L i_1 + \alpha j + \alpha_{iT} \gamma_{Li} i_i] \ , \tag{2a}$$

$$\varphi_i = \beta'_L \gamma_{Li} [\gamma_{Li} i_i + \alpha_i j + \alpha_{iT} \gamma_L i_1] \ , \tag{2b}$$

and

$$2\pi\varphi = 2\pi\varphi_{ext} + \beta'_L [j + \alpha \gamma_L i_1 + \alpha_i \gamma_{Li} i_i] \ . \tag{2c}$$

Here, fluxes through the input circuit and tank circuit inductors are normalized to $\Phi_0 / 2\pi$, and the flux through the SQUID is normalized to $\Phi_0$. We solve these expressions for $j$, $i_1$ and $i_i$ and insert the resulting functions into the differential equations introduced below. For the tank circuit we find $I_T = I_1 + C_T \dot{U}_{CT} + \dot{\Phi}_T / R_A + I_{NA}$ and, for $R_T > 0$, $\dot{U}_{CT} = (\dot{\Phi}_T - U_{CT} + U_{NT}) / R_T C_T$; for $R_T = 0$, this expression reduces to $U_{CT} = \dot{\Phi}_T$. In dimensionless form these equations become

$$i_T = i_1 + \frac{\dot{u}_{cT}}{f_0^2 \beta_L' \gamma_L^2} + \frac{\dot{\phi}_T}{\beta_L' f_0 \gamma_L^2 Q_A} + i_{NA} \quad , \tag{3}$$

$$\dot{u}_{cT} = f_0 Q_0 (\dot{\phi}_T - u_{cT} + u_{NT}) \quad , \qquad (R_T > 0) \tag{4a}$$

and

$$\dot{\phi}_T = u_{cT} \quad . \qquad (R_T = 0) \tag{4b}$$

For the input circuit, in analogy with the tank circuit, we find $C_i \dot{U}_{Ci} = -I_i$ and $\dot{U}_{Ci} = (\dot{\Phi}_i - U_{Ci} + U_{Ni}) / R_i C_i = (\dot{\Phi}_i - U_{Ci}) / R_i C_i + I_{Ni} / C_i$ . In dimensionless form these equations are

$$i_i = -\frac{1}{f_{0i}^2 \beta_L' \gamma_{Li}^2} \dot{u}_{ci} \tag{5}$$

and

$$\dot{u}_{ci} = f_{0i} Q_{0i} (\dot{\phi}_i - u_{ci} + u_{Ni}) \quad . \tag{6}$$

The noise voltage $u_{Ni}$ has a spectral power density $4\Gamma_i R_i / R$, where $\Gamma_i = 2\pi k_B T_i / I_0 \Phi_0$ and $R_i / R = \beta_L' f_{0i} \gamma_{Li}^2 / Q_{0i}$. For the untuned input with $C_i \to \infty$, a short calculation yields

$$\dot{\phi}_i = -\frac{f_{0i} \gamma_{Li}^2 \beta_L'}{Q_{0i}} i_i + u_{Ni} \quad , \tag{7}$$

which replaces Eqs. (5) and (6). Note that the ratio $f_{0i}/Q_{0i}$ is equal to $R_i/2\pi L_i f_c$ and is independent of $C_i$. Finally, the equation for the SQUID loop is

$$\beta_c \ddot{\delta} + \dot{\delta} + \sin \delta + i_N = j .\qquad(8)$$

For the currents, voltages and fluxes in the tank circuit and input circuit, simple scaling relations hold with respect to $\gamma_L$ and $\gamma_{Li}$: for the tank circuit $i_T(\gamma_L) = i_T(\gamma_L = 1)/\gamma_L$, $u_T(\gamma_L) = \gamma_L \cdot u_T(\gamma_L = 1)$ and $\varphi_T(\gamma_L) = \gamma_L \cdot \varphi_T(\gamma_L = 1)$, and for the input circuit $i_i(\gamma_{Li}) = i_i(\gamma_{Li} = 1)/\gamma_{Li}$, $u_i(\gamma_{Li}) = \gamma_{Li} \cdot u_i(\gamma_{Li} = 1)$ and $\varphi_i(\gamma_{Li}) = \gamma_{Li} \cdot \varphi_i(\gamma_{Li} = 1)$. It is thus sufficient to consider the cases $\gamma_L = \gamma_{Li} = 1$.

Equations (2) - (8) contain the variables $\delta$, $\varphi_T$, $u_{cT}$, $\varphi_i$ and $u_{ci}$. The 16 model parameters are $\beta'_L$, $\beta_c$, $i_d$, $\varphi_{ext}$, $f$, $f_0$, $f_{0i}$, $Q_0$, $Q_{0i}$, $Q_A$, $\alpha$, $\alpha_i$, $\alpha_{iT}$, $\Gamma$, $\Gamma_i$, $\Gamma_A$. An *ab initio* optimization of such a large number of parameters is obviously out of reach. Instead, we give an approximate low frequency analysis of the noise temperature and combine it with numerical simulations for some special cases. Details are discussed in Appendix A1. In brief, in the limit $\alpha_{iT} = 0$ and for $\alpha_i^2 \ll 1$ one finds that for a tuned input circuit the dimensionless noise temperature, optimized for $\alpha_i$ and $f_{0i}$, is given by

$$\Gamma_{N,opt} = \frac{2\pi k_B T_{N,opt}}{I_0 \Phi_0} \cong \pi f \sqrt{(\tilde{s}_{vT}\tilde{s}_j - \tilde{s}_{vj}^2)} / v_\varphi .\qquad(9)$$

When the input circuit is at resonance one obtains $\Gamma_{N,res} \approx \pi f \sqrt{\tilde{s}_{vT}\tilde{s}_j} / v_\varphi$. Here, $\tilde{s}_{vT}$, $\tilde{s}_j$ and $\tilde{s}_{vj}$ denote, respectively, the normalized spectral density of the tank circuit noise voltage $v_T$, the circulating noise current $j$ in the SQUID loop and the cross correlation between the Fourier transforms of $v_T$ and $j$, evaluated at $\alpha_i = \alpha_{iT} = 0$. The frequency $f$ is normalized to the junction characteristic frequency $f_c$. We have set $v_\varphi \equiv |\mathrm{d}v_T/\mathrm{d}\varphi_{ext}|$. When the optimal frequency of the input signal is close to the input circuit resonance frequency the optimal value for $Q_{0i}$ is obtained from

$$\alpha_i^2 Q_{0i}\big|_{res} \approx \alpha_{i,opt}^2 Q_{0i}\big|_{res} = 2\pi(\tilde{s}_{vT} / \tilde{s}_j)^{1/2} / v_\varphi \beta_L' \ . \tag{10}$$

Equation (9) is generic for both rf and dc SQUIDs,[3,6] and, to an accuracy of 10-30%, describes the noise temperature of the SQUID coupled to the tuned input circuit. (cf. Appendix A2. The case $\alpha_{iT}$ > 0 is addressed in the appendices.)

## 3. OPTIMIZED NOISE TEMPERATURE

We first optimize Eq. (9) for the fixed values $f_0 = 0.1$, $\alpha_i = \alpha_{iT} = 0$, $\Gamma = 0.025$, $\beta_c = 0$, $Q_0 = 100$ and $Q_A \to \infty$, and find $\Gamma_{N,opt} / f \approx 0.07$. The values of the model parameters we varied to optimize the noise temperature are: $i_d = 0.62$, $f_d = 0.106$, $\alpha = 0.72$, $\varphi_{ext} = 0.28$ and $\beta_L' = 0.68$, yielding $v_\varphi = 1.29$, $\tilde{s}_{vT} = 0.011$, $\tilde{s}_j = 0.15$, $\tilde{s}_{vj} = 0.029$ and $e \approx 0.6$. Figure 2(a) shows (high frequency) spectra for $s_{uT}$ and $\tilde{s}_j$ for the above parameters. For comparison, Fig. 2(b) shows these spectra for the parameters that minimize the noise energy for the above fixed parameters and $\varphi_{ext} = 0.25$: $\alpha = 0.73$, $f_d = 0.1066$, $i_d = 0.37$, $\beta_L' = 1.21$, yielding $e \approx 0.5$. Using these parameters we would have obtained a noise temperature some 30% higher than for case (a). At first sight, the spectra in (a) and (b) appear similar. The essential difference in (b) is a pronounced dip in $s_{uT}$ at f = 0.11 which apparently helps to minimize $e$. This dip is much less pronounced in (a). We further note that the model parameters for both cases are similar, except that $\beta_L'$ is about a factor of 2 lower in case (a).

We now investigate the ratio $\Gamma_{N,opt} / f$, optimized for several model parameters. We first optimize for $i_d$, $f_d$ and $\beta_L'$ under the conditions $\beta_c = 0$, $\varphi_{ext} = 0.25$ and $\Gamma = 0.025$, and study $\Gamma_{N,opt} / f$ as a function of α. Figure 3 shows the resulting plot for the two cases $Q_0 = 100$, $Q_A \to \infty$, with $\Gamma_A = 0$ (preamplifier absent, open squares) and $Q_0 = Q_A = 200$, with $\Gamma_A = 0.25$ (matched case, open circles). Table 1 lists the corresponding model parameters. In the absence of the preamplifier and for $\alpha > 0.4$ we find $\Gamma_{N,opt} / f < 0.1$ ($\Gamma_{N,opt} / \hbar f < 4$); this value increases as $\alpha$ is reduced, for example, rising to about 0.15 for $\alpha = 0.2$.

For $Q_0 = Q_A = 200$ the overall shape of $\Gamma_{N,opt} / f$ vs. $\alpha$ is similar to the case $Q_A \rightarrow \infty$, although the absolute values are roughly a factor of 2 larger. As for the case of the optimized noise energy,[2] the increase of $\Gamma_{N,opt} / f$ for decreasing $\alpha$ scales much less strongly than $\alpha^2$, as is predicted by Eq. (1). The data are more consistent with an $\alpha^1$ scaling, although the numerical data points are not sufficiently accurate to determine the exponent precisely. Nonetheless, the plot shows that a large value of $\alpha$ is highly desirable.

Figure 4 shows $\Gamma_{N,opt} / f$ vs. $\beta'_L$ under various conditions ($Q_0 = 100$ with $Q_A \rightarrow \infty$ and $Q_0 = Q_A = 200$, both cases for $\alpha = 0.2$ and $\alpha$ optimized; $i_d$ and $f_d$ optimized in all cases). Table 2 lists the corresponding model parameters and spectral densities. For all plots $\Gamma = 0.025$, $\Gamma_A = 0.25$, $Q_{eff,0} = 100$, $f_0 = 0.1$, $\beta_c = 0$ and $\varphi_{ext} = 0.25$. In the case $Q_A \rightarrow \infty$, for optimized $\alpha$, $\Gamma_{N,opt} / f$ reaches its minimum of about 0.07 ($\Gamma_{N,opt} / \hbar f \approx 3$) near $\beta'_L = 0.7$. The minimum is very shallow, however, and $\Gamma_{N,opt} / f$ remains below 0.1 over almost the entire range. To achieve such low noise temperatures, values of $\alpha$ above 0.65 are required (Table 2). When we choose $\alpha = 0.2$, the noise temperature increases by a factor of about 2.5. For $Q_0 = Q_A = 200$, $\Gamma_{N,opt} / f$ increases by a factor of roughly 2 for variable $\alpha$, increasing by another factor of 3 when $\alpha$ is fixed at 0.2. In all cases, the minimum values of $\Gamma_{N,opt} / f$ are achieved in the range $0.5 < \beta'_L < 1$ which–in contrast to large values of $\alpha$–are easy to realize experimentally. Note that Eq. (1), which is valid for $\beta'_L \ll 1$, predicts that the minimum noise temperature should be independent of $\beta'_L$. The minimum value of $\Gamma_{N,opt} / \hbar f$ is predicted to be about 2.5, which is close to the value we obtained numerically. For the case $Q_A \rightarrow \infty$, using $T_T/T = 1$ and $Q = Q_0$, and for $\beta'_L \approx 0.7$, we expect the contribution $T_T/T(\alpha^2 Q \beta'_L f_d)$ from the tank circuit to be about 3.5, leading to $\Gamma_{N,opt}/\Gamma f \approx 11$. The value found numerically is 6-7; thus, Eq. (1) overestimates the tank circuit contribution for this case. The same holds for $Q_0 = Q_A = 200$ where, for $\alpha = 0.2$, the minimum numerical values of $\Gamma_{N,opt}/\Gamma f$ are about 13, while from Eq. (1), with $T_T/T = 10$, we predict a value above 40.

Finally, we examine the dependence of $\Gamma_{N,opt}/f$ on $\Gamma$ and $\Gamma_A$ for the optimized parameters $i_d$, $f_d$, $\alpha$ and $\beta'_L$ and fixed parameters $Q_A = Q_0 = 200$ $\alpha_i = \alpha_{iT} = 0$, $f_0 = 0.1$, $\varphi_{ext} = 0.25$ and $\beta_c = 0$. For $\Gamma_A = 0$ the spectral densities and particularly $\Gamma_{N,opt}/f$ vs. $\Gamma$ are approximately linear; Fig. 5(a) shows that $\Gamma_{N,opt}/f \approx 3.2\Gamma$. The optimized parameters are listed in Table 3. It is interesting to note that $\beta'_L$ decreases with increasing $\Gamma$, reaching a value of 0.26 for $\Gamma = 1.6$. Such a decrease can be understood when $\Gamma_{N,opt}/f$ is an increasing function of $\Gamma\beta'_L$, as for the case of the noise energy, where an exponential growth with $\Gamma\beta'_L$ is predicted.[12] This growth counteracts the minimum of $\Gamma_{N,opt}/f$ vs. $\beta'_L$ obtained for fixed $\Gamma$ (Fig. 4). When we increase $\Gamma_A$ for $\Gamma = 0$, the spectral densities follow power laws, with exponents between 1.2 and 1.6 [Fig. 5(b)]. In particular, $\Gamma_{N,opt}/f \approx 250(\Gamma_A/Q_A)^{1.6}$. Here, the parameters $i_d$, $f_d$, $\alpha$ and $\beta'_L$ stay more or less constant, showing that $\Gamma_A$ and $\Gamma$ act quite differently on the noise temperature.

## 4. BROADBAND, FAST READOUTS

So far, we have determined the Fourier component of $s_{uT}$ at the drive frequency by recording a time trace $u_T$ over typically 200 periods of the alternating current drive. Consequently, the bandwidth of our amplitude sensitive detector was typically $5 \cdot 10^{-3} f_d \approx 5 \cdot 10^{-4}$. We next examine cases where this bandwidth is much larger. We also study situations in which the drive frequency becomes comparable with or even larger than the characteristic frequency.

For a first test of such fast readouts we take parameters $Q_0 = 101$, $Q_A = 10100$, $f_0 = 0.1$, $\beta'_L = 1.21$ $\alpha = 0.725$, $i_d = 0.369$, $f_d = 0.1066$, $\varphi_{ext} = 0.25$, $\Gamma = 0.025$ and $\Gamma_A = 0.25$, and use a bandwidth $f_d/8$ (i. e. we take time traces $u_T$ over 8 periods of the current drive). Figure 6 (a) shows the (high frequency) Fourier spectra of $s_{uT}$ and $\tilde{s}_j$ together with the low-frequency correlation functions $\tilde{s}_{vT}$, $\tilde{s}_j$ and $\tilde{s}_{vj}$. As we see, above $f \approx 0.01$ the low-frequency functions increase. This increase actually corresponds to the bump visible in Fig. 1. From these low-frequency spectra, we infer the optimal noise temperature via Eq. (8), the best value of $\alpha_i^2 Q_{0i}$ via Eq. (10) and the noise energy via $e = \pi \tilde{s}_{vT} / 2\Gamma \beta'_L$. The result is shown in Fig. 6(b). It is evident that, at least for this set of parameters, the noise temperature deteriorates above frequencies of about $10^{-3}$, which are an upper limit for minimal noise temperatures.

We next optimize $T_{N,opt}/f$ for $Q_0 = 100$, $Q_A \rightarrow \infty$, $\Gamma = 0.025$, $f_0 = 0.1$ and $\beta_c = 0$, using a bandwidth $f_d/8$. The result is $<T_{N,opt}/f> \approx 0.13$, where the brackets denote averaging over the entire bandwidth. The values of the model parameters we varied to optimize the noise temperature are: $i_d = 0.56$, $f_d = 0.1063$, $\alpha = 0.916$, $\varphi_{ext} = 0.278$ and $\beta'_L = 0.628$, yielding $v_\varphi = 1.31$. Figure 7 shows (a) the corresponding noise power spectra and (b) the inferred values of $\Gamma_{N,opt}/f$, $\Gamma_{N,res}/f$, $\alpha_i^2 Q_{0i}$ and $e$. The low frequency limit of $\Gamma_{N,opt}/f$ ($\approx 0.065$) is comparable to the optimized value using the narrowband readout. Again, above $f \approx 3$ x $10^{-3}$ the ratio $\Gamma_{N,opt}/f$ and the spectral densities increase. The broad bandwidth optimization, which puts an extra weight on the bump near $f = 0.01$, is thus not able to decrease the amplitude of the bump.

In the next step we increase $f_0$ to 0.5. For narrow-band readout, the optimization routine finds $\Gamma_{N,opt}/f = 0.066$ for parameters $f_d = 0.54$, $i_d = 0.894$, $\alpha = 0.53$, $\varphi_{ext} = 0.25$ and $\beta'_L = 0.60$. For the spectral densities we obtain $\tilde{s}_{vT} = 0.092$, $\tilde{s}_j = 0.076$ and $\tilde{s}_{vj} = 0.0036$; in addition $v_\varphi = 3.98$. For these parameters, but now with a bandwidth of $f_d/8$, Fig. 8 shows (a) the spectral densities $s_{uT}$, $\tilde{s}_j$, $\tilde{s}_{vT}$ and $\tilde{s}_{vj}$ and (b) the calculated values of $\Gamma_{N,opt}/f$, $\Gamma_{N,res}/f$, $e$ and $\alpha_i^2 Q_{0i}$. In Fig. 8(a), we note that $\tilde{s}_{vT}$ is now very large and that, on the other hand, $\tilde{s}_{vj}$ becomes small. Consequently, in Fig. 8(b), $\Gamma_{N,opt}/f$ and $\Gamma_{N,res}/f$ are almost indistinguishable. Both functions remain flat up to frequencies of about $10^{-2}$ where the hump, already visible for the low-frequency drives, reappears, although in a less pronounced way.

The above examples show that over a wide regime of drive frequencies and readout bandwidths there are some quantitative differences but no major qualitative changes. With this finding in mind, in Fig. 9, we display the frequency evolution of the noise temperature for several cases. As we see, $\Gamma_{N,opt}/f$ is flat when $\alpha$ is optimized for the case $Q_0 = 100$, $Q_A \rightarrow \infty$ (open squares). For $\alpha = 0.2$ and $Q_0 = 100$, $Q_A \rightarrow \infty$, on the other hand, $\Gamma_{N,opt}/f$ starts to increase for drive frequencies below about 0.3 (black circles). The situation is similar to the case of the optimized noise energy where we found $e$ to increase approximately as $f_d^{-0.3}$. In the case of the noise temperature, $\Gamma_{N,opt}/f$ scales approximately as $f_d^{-0.5}$ for $\alpha = 0.2$ and $f_d < 0.2$ (cf. dotted line which is given by the fit function $0.043/f_0^{1/2}$). For example, for $f_d = 0.01$ the extrapolated value of $\Gamma_{N,opt}/f$ is about 0.4, more than 5 times above the minimum value. We thus see clearly that, for reasonable values of $\alpha$, the drive frequency cannot be too low. For the case of the "noisy preamplifier", with $Q_0 = Q_A = 100$, $\alpha = 0.2$ and $\Gamma_A = 0.25$, for drive frequencies $f_d < 0.2$, $\Gamma_{N,opt}/f$ increases by another factor of 2.5 compared to the isolated case (open circles). We also show with grey circles simulations where, for the case $Q_0 = Q_A = 100$, $\alpha = 0.2$ and $\Gamma_A = 0.25$ we additionally added the preamplifier voltage noise to the output, assuming $\kappa = 0.2$ and $\Gamma_{NA} = 0.05$. The numerical data roughly follow the fit function $0.037/f_0 + 0.1/f_0^{0.5}$. The increase in $\Gamma_{N,opt}/f$, compared to the isolated case, is enormous, showing that at least for these values of $\Gamma_{NA}$ the noise performance of the rf SQUID is limited by the preamplifier noise. In additional simulations (not shown), for $\Gamma_{NA} = 0.05$ we also investigated the case $Q_0 = 100$, $Q_A=10100$ yielding essentially the same result. For $Q_0 = Q_A=200$ we further optimized $\kappa$ and $\alpha$ and investigated the case $Q_{eff,0} = 200$ but did not achieve a significant improvement over the system noise temperature shown by the grey circles in Fig. 9. We thus conclude that, unless the normalized drive frequencies approach unity or $\Gamma_{NA}$ is lower or at most comparable to $\Gamma$ the system noise temperature is dominated by the preamplifier.

## 5. CONCLUSIONS

We have seen that, as for the noise energy,[2] numerical simulations of the rf SQUID based on Langevin equations yield extremely low values of the intrinsic optimal noise temperature. The best values obtained are close to previous analytical estimates[3], although differences occur away from this optimum. For a noise parameter $\Gamma = 0.025$ we found $\Gamma_{N,opt} \approx 0.07 f \approx 3 f \Gamma$ (or $T_{N,opt}/T \approx 3f / f_c$ in absolute units) for $\beta'_L \approx 0.8$ (by contrast, from the analytic theory, which is valid for $\beta'_L \ll 1$, one would have expected a noise temperature that is independent of $\beta'_L$). The linear relation $\Gamma_{N,opt} \approx 3f\Gamma$ holds for $\Gamma$ values up to about 2. With increasing $\Gamma$ the optimal value of $\beta'_L$ decreases, however, reaching a value of about 0.25 at $\Gamma = 1.6$ and $f_d = 0.1f_c$, for example. In the low fluctuation limit the optimum noise temperature of the rf SQUID is a factor of 3 or so lower than for a dc SQUID with an optimized inductance parameter $\beta'_L \approx 1$. This factor increases for larger values of $\Gamma$, since for the dc SQUID the transfer function decreases strongly with $\Gamma$ while for the rf SQUD it remains essentially constant up to $\Gamma \approx 1$.

The drawback with the rf SQUID, however, is that to achieve low values of the noise temperature and noise energy the coupling coefficient $\alpha$ between the SQUID loop and the tank circuit needs to be large: the optimum value is $\alpha \rightarrow 1$. With a more realistic value $\alpha = 0.2$ and for a tank circuit quality factor $Q_0 = 100$, the noise temperature degrades by a factor of about 2.5 for a reduction in drive frequency $f_d$ to a factor of 10 below the junction characteristic frequency $f_c$. The degradation becomes worse for lower values of $f_d$. Assuming $f_c = 100$ GHz and $f_d = 1$ GHz, we find that $\Gamma_{N,opt}$ extrapolates to about $15f\Gamma$, a value slightly higher than can be achieved with a dc SQUID (at low temperatures). With respect to noise temperature, we have not systematically studied the dependence on $Q_0$. However, for the noise energy we found that an increase of $Q_0$ improves $e$ only modestly, by much less than indicated by the predicted proportionality[3] to $1/\alpha^2 Q_0$. A similar dependence is also likely to hold for the noise temperature. Experimentally it will certainly be difficult, although perhaps not impossible, to achieve reasonably large values of $f_d$ and $\alpha$ simultaneously. Another issue regards the preamplifier. We have seen that its noise current and its noise voltage add substantially to the SQUID noise temperature, and in many cases will dominate the system noise temperature.

We give some realistic examples. We assume $f_d = 1$ GHz, $T = 4.2$ K, $f_c = 100$ GHz and $L = 40$ pH. The condition $\beta_L' \approx 0.8$ implies $I_0 \approx 6.5$ μA, leading to $\Gamma \approx 0.027$, close to the value we used for our calculation. For $f_c = 100$ GHz, we find $I_0R \approx 207$ μV, corresponding to R $\approx 32$ Ω. For the optimal noise temperature $3Tf/f_c$, we find $T_{N,opt} / f \approx 0.12$ K per GHz which, for $f = 1$ MHz, results in $T_{N,opt} \approx 120$ μK. The required value of α $\approx 0.6$, however, is difficult to achieve. For the more realistic value $\alpha = 0.2$, we predict a value $T_{N,opt} / f \approx 0.7$ K per GHz (or 0.7 K at 100 MHz; this value is comparable to that achieved with dc SQUIDs at 4.2 K). For comparison, between roughly 10 and 100 GHz, the best cold HEMTs have noise temperatures of about 0.5 K per GHz.

To find the optimal parameters of the tank circuit, for convenience we assume a preamplifier with a noise temperature of 5 K ($\Gamma_{NA} = 0.032$), R$_{opt} = 50$ Ω and $\kappa = 0.2$. With $T_{NA} = S_{IA}^{1/2} S_{UA}^{1/2} / 2k_B$ and $R_{opt} = S_{UA}^{1/2}/S_{IA}^{1/2}$ we obtain a current noise $S_{IA}^{1/2}$ of 1.67 pA/Hz$^{1/2}$ and a voltage noise $S_{UA}^{1/2}$ of 0.083 nV/ Hz$^{1/2}$. For the case $Q_A = 10100$, $Q_0 = 101$ we obtain $R_T \approx R_{opt} / Q_0^2 \kappa \approx 25$ mΩ. With $Q_0 \cdot Q_A = R_A / R_T$ we find $R_A \approx 25$ k$\Omega$. With $T_A \approx T_{NA}Q_A/2\kappa Q_0$ we obtain $T_A \approx 1250$ K, and for $\Gamma \approx$ 0.027 we find $\Gamma_A \approx 8$. The ratio $\Gamma_A/Q_A \approx 8 \times 10^{-4}$ is comparable to the one for the "noisy" amplifier ($\Gamma_A/Q_A = 1.25 \times 10^{-3}$). We thus expect our amplifier to dominate the SQUID noise temperature. (A direct simulation, using the above parameters while varying $f_d$ and $i_d$ confirmed this result, yielding $\Gamma_{n,opt} / f = 4.0$ or $T_{N,opt} \approx 150 Tf/f_c$.) Using $Q_0 = \sqrt{L_T / C_T R_T^2}$ we find $\sqrt{L_T / C_T} \approx 2.5 \ \Omega$ and, with $f_0 \approx 1/2\pi\sqrt{L_T C_T}$ , we obtain $C_T = \sqrt{C_T / L_T} / 2\pi f_0 \approx 70$ pF, $L_T \approx 390$ pH and $\gamma_L^2 \approx 10$.

For the case $Q_0 = Q_A = 200$ a similar calculation leads to $T_A = 25$ K, $\Gamma_A = 0.16$, $R_A = 500 \ \Omega$, $R_T = 12.5$ m$\Omega$, $\sqrt{L_T / C_T} = 2.5 \Omega$, $C_T = 64$ pF , $L_T = 400$ pH and $\gamma_L^2 \approx 10$. Since $\Gamma_A$ is only slightly below the value of our "noisy" preamplifier the resulting noise temperature will be only slightly better than the curves of Fig. 9. Thus, by interpolation, we expect $T_{N,opt} \approx 30 Tf/f_c$ excluding the preamplifier voltage noise, and $T_{N,opt} \approx 150 Tf/f_c$ including it. (A direct simulation confirmed this result, yielding $T_{N,opt} \approx 140 Tf/f_c$ for the latter case.) Our preamplifier thus clearly dominates the system noise temperature.

Some final remarks are in order. It will be challenging – to say the least – to achieve experimentally the ultimate noise temperature of the rf SQUID amplifier predicted by our simulations. There are two over-riding reasons for this limitation. First, in practice, it has proven very difficult[13] to increase the coupling coefficient $\alpha$ to values much above about 0.2, whereas the lowest noise temperature requires $\alpha \rightarrow 1$. It would be of considerable interest to revisit this issue experimentally. Second, the lowest noise temperatures of cooled HEMTs are 1-3 K.[14] Thus, when the rf SQUID is cooled to (say) 20 mK, one requires a power gain of well over 20 dB to make the noise from the HEMT negligible. Such high levels of gain are difficult to achieve. In contrast, the dc SQUID has a higher gain at a frequency of (say) 1 GHz–perhaps 30 dB–so that the noise temperature of the HEMT preamplifier is not a limiting factor when the SQUID is cooled to 20 mK. Thus, although the ultimate noise temperature of the rf SQUID amplifier may, in principle, be comparable to or even lower than that of a dc SQUID amplifier, in practice, this seems unlikely to be realized.

Furthermore, we note that the same issues of parasitic capacitance between the input coil and the SQUID washer apply to both rf and dc SQUIDs, limiting the upper frequency range to about 100 MHz in the conventional mode of operation. In the case of the dc SQUID, this drawback has been successfully overcome by means of the microstrip SQUID amplifier,[15] but to our knowledge, this configuration has not yet been implemented for the rf SQUID.

Finally, the dc SQUID is significantly easier to implement, particularly as a high-frequency amplifier. At the Josephson frequency (say 10-20 GHz), the dc SQUID up-converts the signal parametrically and subsequently down-converts to the original signal frequency–without the need for any external rf signal. For these reasons, the rf SQUID is unlikely to challenge the dc SQUID as an amplifier at frequencies above (say) 100 MHz.

## ACKNOWLEDGMENTS

The authors thank A. I. Braginski, B. Chesca, D. Kinion, M. Mück and Y. Zhang for valuable discussions. We gratefully acknowledge financial support by the Deutsche Forschungsgemeinschaft (R. K. and D. K) and by the Director, Office of Science, Office of Basic Energy Sciences, Materials Sciences and Engineering Division, of the U.S. Department of Energy under Contract No. DE-AC02-05CH11231 (J.C.).

**APPENDIX A1: LOW FREQUENCY ANALYSIS OF THE RF SQUID CIRCUIT**

To analyze the low frequency behavior of the rf SQUID circuit shown in Fig. 1 we follow the strategy of refs. 3 and 6. We investigate flux changes $\delta\Phi$ in the SQUID loop and relate them to the demodulated low frequency tank circuit voltage via $V_T = V_{T0} + (dV_T / d\Phi)\delta\Phi$ [ $v_T = v_{T0} + (dv_T / d\varphi)\delta\varphi$ ]. Here, $v_{T0}$ is the normalized low frequency voltage across the tank circuit at the (optimal) bias point in the absence of the input circuit and the noise induced by the preamplifier. The equivalent low frequency circuit is shown in Fig. A1. The input circuit is described by an impedance $Z_i(\omega)$ [ $Z_i = R_i + i\omega L_i + 1/i\omega C_i$, where $i = (-1)^{1/2}$ )], and contains a voltage noise source $U_i$ with spectral density $4k_B T_i R_i$. We consider this circuit at low frequencies so that it produces negligible noise at the drive frequency of the tank circuit. The SQUID loop with inductance $L$ carries a noise current $J_0$ with spectral density $S_{J0}$, to be determined numerically by solving Eqs. (2) - (8) in the limit $\alpha_i = \alpha_{iT} = 0$.

We describe the preamplifier by an input resistance $R_A$ and two independent noise sources. The first is the short-circuit voltage noise source $U_{NA}$ with spectral density $S_{UA}$ which adds an equivalent noise to the voltage $U_T$ across the tank circuit without backaction on the tank circuit. Its component at the drive frequency is subsequently down-converted to the low frequency output $V_T$. The second preamplifier noise source is the current noise, which induces a high-frequency current noise component, thereby increasing $S_{UT}$ and $S_{VT}$. For the moment, we absorb the high-frequency component into $S_{VT0}$. In our model we assume for simplicity that the preamplifier produces white noise, so that we need to consider a contribution adding noise at low frequencies. The low-frequency component couples into the SQUID loop via $M$ and into the input loop via $M_{iT}$. The low-frequency preamplifier current noise appears as a current noise source with spectral density $S_{IA}$. The low-frequency equivalent circuit of the tank circuit thus consists of a resistor $R_A$, an inductor $L_T$ and a current noise source $I_{NA}$ with spectral density $S_{IA}$. Since there is no low frequency current through the arm consisting of $R_T$ and $C_T$ in Fig. 1, this can be omitted in the low frequency analysis.

Consequently, for $R_A >> \omega L_T$ the low frequency current $I_I$ is given in leading order by $I_{NA}$ ($i_I = i_{NA}$). For the flux change $\delta\varphi$ in the SQUID loop we consider low-frequency contributions coupled to the SQUID by the input circuit and the tank circuit. In dimensionless notation, using $\gamma_L = \gamma_{Li} = 1$, we find $\delta\varphi = \beta'_L(\alpha i_1 + \alpha_i i_i)/2\pi$.

For the noise current in the input circuit we have $I_i = [-U_i + i\omega(M_i J + M_{iT} I_1)]/Z_i$. To lowest order, $J$ is the fluctuating current $J_0$ which is already present in the absence of an input circuit. In dimensionless units, with $z_i = Z_i/i\omega i_i = (1 - f_{0i}^2/f^2) - if_{0i}/fQ_{0i}$ and $\gamma_L = \gamma_{Li} = 1$, we find $i_i = (1/z_i)(iu_i/\beta'_L f + \alpha_i j_0 + \alpha_{iT} i_1)$. In this first order approximation the shift of the input circuit resonance frequency due to the coupling to the SQUID is not taken into account; this is inconvenient for comparisons of the analytic formulas developed below with direct numerical simulations (Appendix A2). We thus go one step further and consider the full flux coupled into the SQUID loop. This leads to $J = J_0 + I_i M_i/L + I_1 M/L$. In dimensionless units, with $\bar{z}_i = z_i - \alpha_i^2$, we find $i_i = [iu_i/\beta'_L f + \alpha_i j_0 + (\alpha_i\alpha + \alpha_{iT})i_1]/\bar{z}_i$.

Assuming that the coupling parameter $\alpha_T$ is small, we keep the first order approximation for $i_I$ and for the flux change in the SQUID loop obtain $\delta\varphi = \alpha i_{NA}\beta'_L/2\pi + \alpha_i(iu_i/f + \alpha_{iT}i_{NA}\beta'_L)/2\pi\bar{z}_i + \beta'_L\alpha_i^2[j_0 + \alpha i_{NA}]/2\pi\bar{z}_i$. For the tank circuit voltage, we find

$$v_T = [v_{T0} + \frac{dv_T/d\varphi}{2\pi}\alpha\beta'_L i_{NA}] + \alpha_i c_i[i\frac{u_i}{f} + \alpha_{iT}\beta'_L i_{NA}] + c_i\alpha_i^2\beta'_L[j_0 + \alpha i_{NA}], \qquad (A1)$$

where $c_1 = (dv_T/d\varphi)/2\pi\bar{z}_i = (dv_T/d\varphi)/2\pi[(1 - f_{0i}^2/f^2 - \alpha_i^2) - if_{0i}/fQ_{0i}]$.

Note that the above expressions take backaction into account only partially. For example, the coupling between the tank circuit and the input loop reduces both $L_i$ and $L_T$ by a factor of $(1 - \alpha_{iT}^2)$. The change in $L_T$, in particular, leads to a detuning of the tank circuit resonance frequency and an increase in the voltage noise power $s_{vT0}$. This effect can be compensated by properly re-adjusting the drive frequency $f_d$, as we shall see in Appendix A2.

We next take the Fourier transform of Eq. (A1) and convert the result into the spectral density

$$s_{vT} = \tilde{s}_{vT} + 4\alpha_i^2 \mid c_1 \mid^2 \cdot \beta_L' (\frac{f_{0i}}{Q_{0i} f^2} \Gamma_i + \frac{\alpha_{iT}^2}{f_0 Q_A} \Gamma_A) + \alpha_i^4 \mid c_1 \mid^2 \beta_L'^2 \tilde{s}_j + \alpha_i^2 \hat{c}_2 \hat{s}_{vj} \ , \qquad \text{(A2)}$$

where $\tilde{s}_{vT} = s_{vT0} + \Gamma_A \cdot v_\varphi^2 \alpha^2 \beta_L' / \pi^2 f_0 Q_A$ and $\tilde{s}_j = s_{j0} + \Gamma_A \cdot 4\alpha^2 / \beta_L' f_0 Q_A$ . In obtaining the expressions for $\tilde{s}_{vT}$ and $\tilde{s}_j$ we assume there is no correlation between $i_{NA}$ and, respectively, $v_T$ and $j_0$. We set $s_{uA} = 4\Gamma_A \gamma_{RA} = 4\Gamma_A \beta_L' f_0 Q_A$ and $s_{ui} = 4\Gamma_i \gamma_{Ri} = 4\Gamma_i \beta_L' f_{0i} / Q_{0i}$ . We further have

$$\hat{c}_2 \hat{s}_{vj} = \beta_L' (c_1 \tilde{j}_{0\omega} \tilde{v}_{T0\omega}^* + c_{1\omega}^* \tilde{j}_{0\omega}^* \tilde{v}_{T0\omega}) \tau = 2\beta_L' \{ \text{Re}(c_1) \tilde{s}_{vj} + \text{Im}(c_1) \breve{s}_{vj} \}$$

$$= \beta_L' v_\varphi \{ \text{Re}(\bar{z}_i) \tilde{s}_{vj} - \text{Im}(\bar{z}_i) \breve{s}_{vj} \} / \pi \mid \bar{z}_i \mid^2 ,$$

where

$$\tilde{s}_{vj} = \text{Re}(\tilde{j}_{0\omega}) \text{Re}(\tilde{v}_{T0\omega}) + \text{Im}(\tilde{j}_{0\omega}) \text{Im}(\tilde{v}_{T0\omega}) \approx s_{vj} + \Gamma_A 2 v_\varphi \alpha^2 / \pi f_0 Q_A$$

and

$$\breve{s}_{vj} = \text{Re}(\tilde{j}_{0\omega}) \text{Im}(\tilde{v}_{T0\omega}) - \text{Im}(\tilde{j}_{0\omega}) \text{Re}(\tilde{v}_{T0\omega}) \approx \text{Re}(j_{0\omega}) \text{Im}(v_{T0\omega}) - \text{Im}(j_{0\omega}) \text{Re}(v_{T0\omega}) .$$

The asterisk denotes the complex conjugate, the tilde represents noise contributions for $\Gamma_A > 0$ and the subscript $\omega$ is the frequency component of the Fourier transforms. Again, we have assumed that the Fourier components of $u_A$ are not correlated with the other terms.

In the expressions $\tilde{s}_{vT}$, $\tilde{s}_j$, $\tilde{s}_{vj}$ and $\breve{s}_{vj}$, apart from the low frequency noise contributions of the resistor $R_A$, we should also make the high-frequency components of the preamplifier current and voltage noise more explicit. We do so by using the more general forms $\tilde{s}_{vT} = s_{vT0} + a\Gamma_A / Q_A$ , $\tilde{s}_j = s_{j0} + b\Gamma_A / Q_A$ and $\tilde{s}_{vj} = s_{vj} + c\Gamma_A / Q_A$ , with coefficients $a$, $b$, $c$ to be determined numerically. In Appendix A2 we see that in general the above correlation functions do not depend linearly on $\Gamma_A/Q_A$, The dependence is somewhat stronger than linear. Although it would be helpful to have analytic expressions for these coefficients, we were not able to obtain them. As a result, in general, $\tilde{s}_{vT}$, $\tilde{s}_j$, $\tilde{s}_{vj}$ and $\bar{s}_{vj}$ must be determined from the low frequency tank circuit voltage and the SQUID circulating current by numerically solving Eqs. (2) to (8) in the limit $\alpha_i = \alpha_{iT} = 0$, that is, in the absence of the input loop.

The noise floor in Eq. (A2) for $\Gamma_i = 0$ is

$$s_{vT,F} = \tilde{s}_{vT} + 4\alpha_i^2 \Gamma_A \mid c_1 \mid^2 \cdot \beta_L' \alpha_{iT}^2 / f_0 Q_A + \alpha_i^4 \mid c_1 \mid^2 \beta_L'^2 \tilde{s}_j + \alpha_i^2 \hat{c}_2 \hat{s}_{vj},$$

and we thus have

$$s_{vT} = s_{vT,F} + 4\alpha_i^2 \Gamma_i \mid c_1 \mid^2 \beta_L' f_{0i} / Q_{0i} f^2.$$

The normalized noise temperature, which we find via $s_{vT} = 2s_{vT,F}$ or $s_{vT,F} = 4\Gamma_N \alpha_i^2 \beta_L' f_{0i} \mid c_1 \mid^2 / Q_{0i} f^2$, is

$$\Gamma_N = \frac{\tilde{s}_{vT} + 4\alpha_i^2 \mid c_1 \mid^2 \cdot \beta_L' \dfrac{\alpha_{iT}^2}{f_0 Q_A} \Gamma_A + \alpha_i^4 \mid c_1 \mid^2 \beta_L'^2 \tilde{s}_j + \alpha_i^2 \hat{c}_2 \hat{s}_{vj}}{4 \mid c_1 \mid^2 \alpha_i^2 \beta_L' f_{0i} / f^2 Q_{0i}} \qquad (A3)$$

or

$$\Gamma_N = \frac{f^2 Q_{0i} \tilde{s}_{vT}}{4 \mid c_1 \mid^2 \alpha_i^2 \beta_L' f_{0i}} + \alpha_i^2 \frac{f^2 Q_{0i} \beta_L' \tilde{s}_j}{4 f_{0i}} + \alpha_{iT}^2 \frac{f^2 Q_{0i} \Gamma_A}{f_{0i} f_0 Q_A} + \frac{f^2 Q_{0i} \hat{c}_2 \hat{s}_{vj}}{4 \mid c_1 \mid^2 \beta_L' f_{0i}} \quad . \qquad (A4)$$

Inserting the above expressions for $\mid c_1 \mid^2$ and $\hat{c}_2 \hat{s}_{vj}$ we have

$$\Gamma_N = \frac{\pi^2 f^2 Q_{0i} \tilde{s}_{vT}}{v_\varphi^2 \beta_L' f_{0i}} \frac{\mid \bar{z}_i \mid^2}{\alpha_i^2} + \alpha_i^2 \frac{f^2 Q_{0i} \beta_L' \tilde{s}_j}{4 f_{0i}} + \alpha_{iT}^2 \frac{f^2 Q_{0i} \Gamma_A}{f_{0i} f_0 Q_A} + \frac{\pi f^2 Q_{0i} \{ \mathrm{Re}(\bar{z}_i) \tilde{s}_{vj} - \mathrm{Im}(\bar{z}_i) \bar{s}_{vj} \}}{v_\varphi f_{0i}}, \quad (A5)$$

with $\bar{z}_i = (1 - f_{0i}^2 / f^2 - \alpha_i^2) - i f_{0i} / f Q_{0i}$.

This expression must now be minimized first with respect to $\alpha_i$ and later on with respect to $f_{0i}$. In terms of $\alpha_i$ the expression to be minimized is of the form

$$\frac{\pi^2 \tilde{s}_{vT}}{v_\varphi^2 \beta_L'} \frac{1}{\alpha_i^2} [(1 - f_{0i}^2 / f^2 - \alpha_i^2)^2 + \frac{f_{0i}^2}{f^2 Q_{0i}^2}] + \alpha_i^2 (\frac{\beta_L' \tilde{s}_j}{4} - \frac{\pi \tilde{s}_{vj}}{v_\varphi}) \quad . \tag{A6}$$

Although the minimization can be performed analytically, the resulting expression is quite complicated and we thus prefer an approximate treatment under the assumption that $\alpha_i^2$ is small, so that $\bar{z}_i \approx z_i$. We then minimize $\pi^2 f^2 Q_{0i} \tilde{s}_{vT} | z_i |^2 / v_\varphi^2 \beta_L' \alpha_i^2 + \alpha_i^2 f^2 Q_{0i} \beta_L' \tilde{s}_j / 4$, and find

$$\alpha_{i,opt}^4 = \frac{4\pi^2 \tilde{s}_{vT} | z_i |^2}{v_\varphi^2 \beta_L'^2 \tilde{s}_j} \quad . \tag{A7}$$

Inserting Eq. (A7) into Eq. (A5), we obtain

$$\Gamma_{N,opt.\alpha i} = \frac{\pi f^2 Q_{0i}}{v_\varphi f_{0i}} (| \bar{z}_i | \sqrt{\tilde{s}_{vT} \tilde{s}_j} + \{ \mathrm{Re}(\bar{z}_i) \tilde{s}_{vj} - \mathrm{Im}(\bar{z}_i) \breve{s}_{vj} \}) + \alpha_{iT}^2 \frac{f^2 Q_{0i} \Gamma_A}{f_{0i} f_0 Q_A} \quad . \tag{A8}$$

Note that in Eq. (A8) $\alpha_{i,opt}$ is also contained in $\bar{z}_i$. Inserting the expression for $\bar{z}_i$ into Eq. (A8) we obtain

$$\Gamma_{N,opt,\alpha i} = \frac{f^2 Q_{0i}}{f_{0i}} \frac{\pi}{v_\varphi} \cdot \left\{ \sqrt{(1 - \alpha_{i,opt}^2 - \frac{f_{0i}^2}{f^2})^2 + \frac{f_{0i}^2}{f^2 Q_{0i}^2}} \cdot \sqrt{\tilde{s}_{vT}\,\tilde{s}_j} + (1 - \alpha_{i,opt}^2 - \frac{f_{0i}^2}{f^2})\tilde{s}_{vj} + \frac{f_{0i}}{f Q_{0i}} \bar{s}_{vj} \right\}$$

$$+ \alpha_{iT}^2 \frac{f^2 Q_{0i} \Gamma_A}{f_{0i} f_0 \, Q_A}. \qquad (A9)$$

Under the resonance condition $f = f_{0i} / \sqrt{1 - \alpha_{i,opt}^2}$ , this expression becomes

$$\Gamma_{N,res} = \frac{\pi f}{v_\varphi} \cdot \left\{ \sqrt{\tilde{s}_{vT}\,\tilde{s}_j} + \bar{s}_{vj} \right\} + \frac{Q_{0i}}{Q_A} \frac{f_{0i}}{f_0} \frac{\alpha_{iT}^2}{\sqrt{1 - \alpha_{i,opt}^2}} \Gamma_A \quad . \qquad (A10)$$

Equation (A10) can also be expressed as

$$\Gamma_{N,res} = \pi f \cdot \left( \sqrt{2\tilde{e} \frac{(\tilde{s}_j / \Gamma)\beta_L'}{\pi}} + \frac{\tilde{s}_{vj}}{v_\varphi \Gamma} \right) \cdot \Gamma + \frac{Q_{0i}}{Q_A} \frac{f_{0i}}{f_0} \frac{\alpha_{iT}^2}{\sqrt{1 - \alpha_{i,opt}^2}} \Gamma_A \quad , \qquad (A11)$$

where $\tilde{e} = \pi \tilde{s}_{vT} / 2 v_\varphi^2 \Gamma \beta_L'$ . In numerical simulations (Appendix A2), we find $\bar{s}_{vj}$ to be very small and neglect it. From Eq. (A7), at resonance and assuming $\alpha_i^2 \ll 1$, we further obtain

$$\alpha_{i,opt}^2 Q_{0i} \big|_{res} = 2\pi (\tilde{s}_{vT} / \tilde{s}_j)^{1/2} / v_\varphi \beta_L' = [8\pi \tilde{e} / \beta_L'(\tilde{s}_j / \Gamma)]^{1/2} .$$

Using $\Gamma_A = 0$ and model parameters that lead to a small noise energy ( $\beta_L' \approx 1.2$ , $\tilde{e} \approx 0.5$ , $\tilde{s}_j / \Gamma$ $\approx 4$), we obtain $\alpha_{i,opt}^2 Q_{0i} \approx 1.6$ and, neglecting $\bar{s}_{vj}$ , $\Gamma_{N,res} / \Gamma \approx 3.9 f$ . This value is more than a factor of 4 lower than the corresponding noise temperature of a dc SQUID for $\beta_L = 1$ ( $\Gamma_{N,res} / \Gamma \approx 18 f$ ).[16] Even lower values can be obtained by re-optimizing all model parameters. We address this in Sec. 3.

Next, for $\alpha_{iT} = 0$ we optimize Eq. (A9) with respect to $f_{0i}$. We assume $\alpha_i^2 \ll 1$ and neglect $\bar{s}_{vj}$. Then, with $x = f_{0i}/f$ and $y = (1-x^2)/x$, from Eq. (A9) we find $\Gamma_{N,opt,\alpha}(\alpha_{iT} = 0) \approx \pi f \cdot (\sqrt{Q_{0i}^2 y^2 + 1} \cdot \sqrt{\tilde{s}_{vT}\tilde{s}_j} + Q_{0i}\tilde{s}_{vj}y)/v_\varphi$. Note that for $\tilde{s}_{vj} > 0$, $y$ must be negative to obtain a minimum in $\Gamma_{N,opt,\alpha}(\alpha_{iT} = 0)$. Optimizing for $y$ yields $y_{opt}^2 = \tilde{s}_{vj}^2/Q_{0i}^2(\tilde{s}_v\tilde{s}_j - \tilde{s}_{vj}^2)$, from which we find the optimal noise temperature

$$\Gamma_{N,opt}(\alpha_{iT} = 0) = \frac{\pi f}{v_\varphi} \cdot (\tilde{s}_{vT}\tilde{s}_j - \tilde{s}_{vj}^2)^{1/2} \quad . \tag{A12}$$

Equation (A12) is precisely the result found for the dc SQUID.[6]

The optimal frequency $f_{0i}$ is obtained via $y_{opt}$ by solving

$$\left[\frac{1 - \alpha_{i,opt}^2 - (f_{0i}/f)^2}{f_{0i}/f}\right]_{opt} = -\frac{\tilde{s}_{vj}}{Q_{0i}^2\sqrt{(\tilde{s}_{vT}\tilde{s}_j - \tilde{s}_{vj}^2)}} \quad . \tag{A13}$$

In many cases we are interested in signal frequencies near $f_{0i}$, that is, for $f_{0i}/f \approx 1$. Under these circumstances, we find $\alpha_i^2 Q_{0i}|_{opt} \approx \alpha_i^2 Q_{0i}|_{res}$.

One could perform a similar optimization to that above including the $\alpha_{iT}$ coupling term. The resulting expression would be somewhat complicated, and we have elected not to pursue this issue.

We now address the preamplifier input voltage noise $u_{NA}$, which adds a voltage spectral density $s_{uA}$ to the output signal at the drive frequency. After down-conversion, this results in an increased low-frequency noise voltage $v_{NA}$ across the tank circuit. With $\kappa = R_{opt}[R_A^{-1} + (Q_0^2 R_T)^{-1}] = \gamma_L = 1$, $Q_{eff,0} = 100$, $f_0 = 0.1$ and $\beta_L = 0.8$, this contribution can be estimated as $s_{uA} = 4\kappa^2 Q_{eff,0}^2\beta_L'\gamma_L^2 f_0\Gamma_A/Q_A$, leading to a contribution of about 3000 to the coefficient $a$. For practical preamplifiers, this term may dominate $\tilde{s}_{vT}$ unless $\kappa$ is small. In addition, the low-frequency voltage noise is further increased by the low-frequency voltage noise of the preamplifier, a contribution which we ignore here. There is no direct contribution of preamplifier voltage noise to the coefficient $b$. Due to the readjustment of various model parameters during the optimization of the noise temperature $\Gamma_{N,opt}$, however, $b$ depends indirectly on $s_{uA}$. We have seen in simulations that, when $s_{uA}$ is increased for fixed $\Gamma_A / Q_A$, $\tilde{s}_j$ systematically decreases (by a factor of more than 2 when $\kappa$ is varied between 0 and 1).

One often considers the preamplifier noise temperature to be fixed. In this case, $s_{uA} = 2\kappa Q_{eff,0} \beta'_L \gamma_L^2 f_0 \Gamma_{NA}$ scales with $\kappa$ while $s_{iA} = 2\Gamma_{NA} / \kappa Q_{eff,0} \beta'_L \gamma_L^2 f_0$ scales as $1/\kappa$. Thus, the noise temperature has an optimal value. In simulations using $f_0 = 0.5$, 0.1 and 0.015 and $Q_A = Q_0 = 200$, $\Gamma_{NA} = 0.05$, $\alpha = 0.2$, and varying $f_0$, $i_d$ and $\beta'_L$, we found a quite flat dependence of $\Gamma_{N,opt}$ on $\kappa$ in the range $0.1 < \kappa < 1$. We further have $R_{opt} / R = \kappa Q_{eff,0} \beta'_L \gamma_L^2 f_0$. Assuming $R_{opt}$, $R$, $Q_{eff,0}$, $f_0$ and $\beta'_L$ to be fixed, we find $\gamma_L^2 \propto \kappa^{-1}$. Thus, $\gamma_L$, which otherwise appears as a scaling parameter, is also fixed once a certain value of $\kappa$ is chosen.

# APPENDIX A2: COMPARISON OF ANALYTIC RESULTS AND NUMERICAL SIMULATIONS

We now return to numerical calculations and check the validity of Eq. (A2). We first consider the parameters $\alpha_{iT} = 0$ $\Gamma_A = 0$, $Q_0 = 101$ and $Q_A = 10100$ (to keep $Q_{eff0} = 100$). We further use our "standard" parameters $f_0 = 0.1$, $\beta'_L = 1.21$ $\alpha = 0.725$, $i_d = 0.369$, $f_d = 0.1066$, $\varphi_{ext} = 0.25$ and $\Gamma = \Gamma_T = 0.025$. In the limit $\alpha_i = 0$ we obtain $v_\varphi = 1.092$ and $e = 0.564$. For the spectral densities we find $s_{vT0} = 0.0132$, $s_j = 0.119$, $s_{vj} = 0.022$ and $\tilde{s}_{vj} \approx 2 \cdot 10^{-4}$; the value of $\tilde{s}_{vj}$ is an upper limit. All spectral densities have a white power spectrum in the frequency range 5 x $10^{-7}$ to 5 x $10^{-4}$ used for the calculation. We use these numbers, together with $\alpha_i = 0.2$, $Q_{0i} = 50$ (which is close to the optimal $Q_{0i,res} \approx 40$) and $f_{0i} = 10^{-4}$, to solve Eq. (A2) for two values of $\Gamma_i$ (0 and 1 x $10^{-5}$). These results are shown in Fig. A2. In terms of the voltage noise spectral density (Fig. A2, inset), numerical simulations and Eq. (A2) are in good agreement, although the numerical values of $s_{vT}$ are somewhat lower than the analytical results. The main graph of Fig. A2 shows the signal-to-noise ratio (SNR). The numerical and analytical SNRs are close to each other. In particular, for the numerical SNR, the peak value is close to 1, showing that $\Gamma_i \approx 1$ x $10^{-5}$ at $f \approx 10^{-4}$ is the noise temperature for the above parameters. For the optimal noise temperature, from Eq. (A12) we would expect a value of about 0.95 x $10^{-5}$, whereas Eq. (A10) yields 1.14 x $10^{-5}$ for the noise temperature on resonance. We thus see that, in terms of noise temperature, our analytical and numerical results are in good agreement for the parameter set used here.

For the above model parameters we next study the case $\Gamma_A > 0$. We choose $\Gamma_A = 0.25$, that is, $10\Gamma$. Under the condition $\alpha_i = \alpha_{iT} = 0$ we obtain $\tilde{s}_{vT} = 0.0138$, $\tilde{s}_j = 0.1255$, $\tilde{s}_{vj} = 0.0224$, and $\bar{s}_{vj} = 6.4$ x $10^{-5}$. These numbers are very close to those for $\Gamma_A = 0$, and we use $Q_{0i} = 50$ as above to calculate $s_{vT}(f)$ and the SNR for $\alpha_i = 0.2$. The results for the SNR for $\alpha_i = 0.2$ and $\alpha_{iT} = 0$ are shown in Fig. A3(a). The numerical and analytical SNRs agree very well. By contrast, for $\alpha_i = 0.2$ and $\alpha_{iT} = 0.2$, we find that the numerical curve $s_{vT}(f)$ has a white contribution that is a factor of about 2 larger than we predict from Eq. (A2) using the value of $\tilde{s}_{vT}$ obtained in the limit $\alpha_{iT} = 0$. As a consequence, the SNR predicted from Eq. (A2) is a factor of roughly 2 larger than the SNR obtained numerically [Fig. A3 (b), grey dashed line]. This difference is likely due to the change in the tank circuit inductance by a factor $1 - \alpha_{iT}^2$, leading to an increase in $f_0$ by a factor $\sqrt{1 - \alpha_{iT}^2}$. To show this, in the simulations we re-adjusted the drive frequency by 2% to $f_d = 0.1088$. The white background in $s_{vT}$ indeed decreases and the re-calculated SNR [Fig. A3 (b), solid grey line] becomes much closer to the analytical curve [Fig. A3 (b), solid black line].

We next study lower values of $Q_A$. Current fluctuations in the resistor $R_A$ are a function of $\Gamma_A / Q_A$. To first order, for a fixed value of $Q_{eff,0}$, we thus expect the spectral densities $\tilde{s}_{vT}$, $\tilde{s}_j$ and $\tilde{s}_{vj}$ to be inversely proportional to $Q_A$, as indicated by the expressions $\tilde{s}_{vT} \approx s_{vT0} + a\Gamma_A / Q_A$, $\tilde{s}_j \approx s_{j0} + b\Gamma_A / Q_A$ and $\tilde{s}_{vj} \approx s_{vj} + c\Gamma_A / Q_A$ given in Appendix A1. On the other hand, there will be down-converted high frequency contributions to the noise power. Also, at least for large fluctuations, the circuit may become detuned, leading to a non-linear dependence of the spectral densities on $Q_A^{-1}$ and $\Gamma_A$. For $Q_{eff 0} = 100$, $\Gamma = 0$, and $\alpha_i = \alpha_{iT} = 0$, Fig. A4 shows $\tilde{s}_{vT}$, $\tilde{s}_j$, $\tilde{s}_{vj}$, $v_\varphi$ and the calculated expression $\pi[\tilde{s}_{vT}\tilde{s}_j - \tilde{s}_{vj}^2]^{1/2} / v_\varphi = \Gamma_{N,opt} / f$ as functions of $\Gamma_A / Q_A$ for fixed $\Gamma_A = 0.25$. For the other parameters we set $f_0 = 0.1$, $\varphi_{ext} = 0.25$, $\beta_c = 0$, $f_d = 0.108$, $i_d = 0.419$, $\alpha = 0.763$ and $\beta'_L = 1.216$ (these values actually minimize the noise energy for $\Gamma = 0.025$). The spectral densities scale almost as $Q_A^{-1}$, although in the regime we investigated the exponent is actually about -1.05. Next, for $Q_0 = Q_A = 200$, $\Gamma = 0.025$ and the other model parameters in Fig. A4, we investigate the spectral densities as functions of $\Gamma$ and $\Gamma_A$. Figure A5 (a) shows the dependence on $\Gamma < 2$ for $\Gamma_A = 0$. For larger values, $v_\varphi$ and thus $\Gamma_{N,opt}$ start to degrade strongly. While in the regime investigated $v_\varphi$ is essentially constant (it actually increases slightly), the spectral densities for $\Gamma < 0.3$ roughly follow a power law with an exponent close to 1.2. In the figure, we show this as a dotted line for the resulting expression for $\Gamma_{N,opt} / f$, which scales as $7\Gamma^{1.2}$ over the entire regime. Figure A5 (b) shows the spectral densities vs. $\Gamma_A/Q_A$ for $\Gamma = 0$. Note that $\Gamma_A/Q_A$ runs over much larger values than in Fig. A4, where we varied $Q_A$. The plot shows that $\tilde{s}_{vT}$, $\tilde{s}_j$ and $\tilde{s}_{vj}$ increase even more strongly than a power law. Nonetheless the calculated function $\Gamma_{N,opt}/f$ is almost linear in the plot, scaling roughly as $200(\Gamma_A/Q_A)^{1.4}$ (dotted line). Further, apart from the nonlinear increase, the values of the correlation functions are significantly larger than one would expect from the low frequency noise alone. For example, for $\Gamma_A = 1$ the low-frequency noise contributions yield $\tilde{s}_{vT} \approx 4 \cdot 10^{-3}$, $\tilde{s}_j \approx 0.096$ and $\tilde{s}_{vj} \approx 0.0194$. The simulated values are, respectively, factors of 4.3, 3.3 and 3.25 larger. To check whether or not the nonlinear dependence of the correlation functions on $\Gamma$ and $\Gamma_A$ arises from detuned parameters $i_d$, $f_d$ etc. we also performed simulations where we varied $i_d, f_d, \alpha$ and $\beta'_L$ to minimize $\Gamma_{N,opt} / f = \pi \sqrt{\tilde{s}_{vT} \tilde{s}_j - \tilde{s}_{vj}^2} / v_\varphi$. Details are given in Sec. 3. In brief, although $\Gamma_{N,opt} / f$ decreases by a factor of about 2 compared to the case of Fig. A5, for $\Gamma = 0$, the spectral densities as a function of $\Gamma_A$ again follow power laws, with exponents between 1.2 and 1.6 [see also Fig 6 (b)]; in particular, $\Gamma_{N,opt} / f \approx 250(\Gamma_A / Q_A)^{1.6}$. As a function of $\Gamma$ for $\Gamma_A = 0$, on the other hand, $\Gamma_{N,opt} / f$ turns out to be nearly linear, $\Gamma_{N,opt} / f \approx 3\Gamma$ [Fig. 6 (a)]. Finally, Fig. A5 (c) shows the spectral densities vs. $\Gamma$ for $\Gamma_A = 1$. For $\Gamma_{N,opt} / f$, by linear superposition we expect $\Gamma_N / f \approx 0.12 + 7\Gamma^{1.2}$. The fit shown as a dotted line is $\Gamma_N / f \approx 0.16 + 7\Gamma^{1.2}$, which is not too far from this expectation.

We next study the SNR of a circuit with $Q_A = Q_0 = 200$. Model parameters are those used above. For $\Gamma_A = 0.25$ and $\alpha_i = \alpha_{iT} = 0$ the simulations yield $\tilde{s}_{vT} = 0.0154$, $\tilde{s}_j = 0.1814$, $\tilde{s}_{vj} = 0.0334$, $\tilde{s}_{vj} = 3 \cdot 10^{-5}$ and $v_\varphi = 1.036$. Inserting these values into the expression for the optimal noise temperature yields $\Gamma_{N,opt}(\alpha_{iT} = 0) \approx 0.125 f_{0i}$, obtained at $\alpha_i^2 Q_{0i} \approx 1.45$, or $Q_{0i} \approx 35$ for $\alpha_i = 0.2$. Under resonance conditions we find $\Gamma_{N,res} \approx 0.161 f_{0i}$. Figure A6 (a) shows simulation results for $Q_{0i} = 30$, $f_{0i} = 10^{-4}$, $\alpha_i = 0.2$, $\alpha_{iT} = 0$ and two values of $\Gamma_i$ (0 and 1.5 x $10^{-5}$). Again, the agreement of the SNR with the analytical curve is very good. Finally, Fig. A6 (b) shows the SNR for the case $\alpha_{iT} = 0.2$ and $Q_{0i} = 50$. Here, both with and without a re-adjustment of $f_d$, the numerical SNR remains about 30 % below the analytical curve, indicating that for these parameters the analytical expressions somewhat underestimate the noise temperature.

To conclude this appendix, we give an example for an untuned input circuit. One can obtain this case from the tuned circuit by setting $f_{0i}$ and $Q_{0i} = 0$ while keeping the ratio $r = f_{0i} / Q_{0i}$ (which corresponds to the rolloff frequency of the input circuit) finite. We thus have $\bar{z}_i = 1 - \alpha_i^2 - ir/f$ .

Equation (A2) becomes

$$s_{vT} = \tilde{s}_{vT} + \frac{\alpha_i^2 v_\varphi^2 \beta_L'}{\pi^2 \mid \bar{z}_i \mid^2} \cdot \frac{r}{f^2} \, \Gamma_i + \alpha_i^4 \frac{v_\varphi^2 \beta_L'^2}{4\pi^2 \mid \bar{z}_i \mid^2} \, \tilde{s}_j + \alpha_{iT}^2 \frac{\alpha_i^2 v_\varphi^2 \beta_L' \Gamma_A}{\pi^2 f_0 Q_A \mid \bar{z}_i \mid^2} + \alpha_i^2 \frac{\beta_L' v_\varphi}{\pi} \{ \frac{\mathrm{Re}(\bar{z}_i)}{\mid \bar{z}_i \mid^2} \tilde{s}_{vj} - \frac{\mathrm{Im}(\bar{z}_i)}{\mid \bar{z}_i \mid^2} \bar{s}_{vj} \} .$$

$$(A14)$$

For the noise temperature, Eq. (A5) yields

$$\Gamma_N = \frac{f^2}{r} \left[ \frac{\pi^2 \tilde{s}_{vT}}{v_\varphi^2 \beta_L'} \frac{\mid \bar{z}_i \mid^2}{\alpha_i^2} + \alpha_i^2 \frac{\beta_L' \tilde{s}_j}{4} + \alpha_{iT}^2 \frac{\Gamma_A}{f_0 Q_A} + \frac{\pi\{(1-\alpha_i^2)\tilde{s}_{vj} + r\bar{s}_{vj}/f\}}{v_\varphi} \right] . \quad (A15)$$

For the simulations we chose $r = 10^{-4}$ together with the parameters $Q_0 = 101$, $Q_A = 10100$, $\alpha_i = 0.2$, $\alpha_{iT} = 0$, $f_0 = 0.1$, $\beta_L' = 1.21$ $\alpha = 0.725$, $i_d = 0.369$, $f_d = 0.1066$, $\varphi_{ext} = 0.25$, $\Gamma = 0.025$, $v_\varphi = 1.092$, $\tilde{s}_{vT} = 0.0132$, $\tilde{s}_j = 0.119$, $\tilde{s}_{vj} = 0.022$ and $\bar{s}_{vj} \approx 2 \cdot 10^{-4}$. The resulting curves $s_{vT}(f)$ for $\Gamma_i = 0$ and $5 \times 10^{-4}$ are shown in Fig. A7(a). For both values of $\Gamma_i$, the numerical data lie below the analytical curves. As a consequence, the analytical SNR is about 30% above the numerical SNR [Fig. A7(b)], and consequently the noise temperature is underestimated by about the same amount.

We see that the analytical equations for the (optimized) noise temperature are in reasonable, although not perfect, agreement with the numerical results. Accepting systematic errors in the range 10-30%, one can go a step further and optimize the expression $\Gamma_{N,opt}/f = \pi\sqrt{(\tilde{s}_{vT}\tilde{s}_j - \tilde{s}_{vj}^2)}/v_\varphi$, as is done in Secs. 3 and 4.

**APPENDIX A3: SYMBOLS USED IN THIS PAPER (IN ADDITION TO THE SYMBOLS OF REFERENCE [2])**

$C_i$: capacitance of input circuit

$\hat{c}_2 \hat{s}_{vj} = \beta'_L v_\varphi \{ \text{Re}(\bar{z}_i) \tilde{s}_{vj} - \text{Im}(\bar{z}_i) \bar{s}_{vj} \} / \pi \mid \bar{z}_i \mid^2$: expression appearing in Eq. (A2)

$f$: frequency normalized to $f_c$

$f_{0i} = \Phi_0 / 2\pi I_0 R \sqrt{L_i C_i}$ : normalized resonance frequency of input circuit

$i_i = I_i/I_0$: normalized current through inductor $L_i$

$I_i$: current through inductor $L_i$

$I_{NA}$: noise current of preamplifier, assumed to have a white spectral density $4 k_B T_A / R_A$

$i_{NA}$: normalized noise current of preamplifier, assumed to have a white spectral density

$J_0$: low frequency circulating current in SQUID loop in the limit $\alpha_i = \alpha_{iT} = 0$, $Q_A \rightarrow \infty$

$j_0$: normalized low frequency circulating current in SQUID loop in the limit $\alpha_i = \alpha_{iT} = 0$,

$\quad Q_A \rightarrow \infty$

$j_{0\omega}$: Fourier component at frequency $\omega$ of normalized low frequency circulating current in SQUID loop in the limit $\alpha_i = \alpha_{iT} = 0$, $Q_A \rightarrow \infty$

$\tilde{j}_{0\omega}$: Fourier component at frequency $\omega$ of normalized low frequency circulating current in SQUID loop in the limit $\alpha_i = \alpha_{iT} = 0$, $Q_A < \infty$, $\Gamma_A > 0$

$L_i$: Inductance of input circuit

$M_i$: mutual inductance between input circuit and SQUID

$M_{iT}$: mutual inductance between input circuit and tank circiut

$Q_A = \sqrt{R_A^2 C_T / L_T}$

$Q_{eff,0} = Q_0 Q_A / (Q_0 + Q_A)$ : effective quality factor of tank circuit unloaded by SQUID

$Q_{0i} = (L_i / C_i)^{1/2} / R_i$  quality factor of input circuit unloaded by SQUID

$R_A$:  effective input resistance of preamplifier

$R_i$:  resistance of input circuit

$R_{opt} = S_{VA}^{1/2} / S_{IA}^{1/2}$:  optimal source resistance of preamplifier at drive frequency

$r = f_{0i}/Q_{0i}$: rolloff frequency of untuned input circuit

$S_{IA} = 4k_B T_A / R_A$:  spectral density of preamplifier noise current, assumed to be white

$s_{iA} = 4\Gamma_A / Q_A \beta'_L f_0 \gamma_L^2$:  normalized spectral density of preamplifier noise current, assumed to be white

$S_J$:  spectral density of circulating current noise in SQUID

$s_j = S_J / I_0 \Phi_0 / 2\pi R$:  normalized spectral density (at all frequencies) of circulating current noise in SQUID

$s_{j0}$:  normalized spectral density (at all frequencies) of circulating current $j$ in the absence of input circuit and noise due to $R_A$ (i. e. $\alpha_i = \alpha_{iT} = 0$, $Q_A \rightarrow \infty$ )

$\tilde{s}_j$:  normalized spectral density (at all frequencies) of circulating current $j$ in SQUID in the absence of input circuit ($\alpha_i = \alpha_{iT} = 0$); $\tilde{s}_j \approx s_{j0} + b\Gamma_A / Q_A$; $b$ is a numerical coefficient

$S_{UA}$:  spectral density of short circuit voltage noise $U_{NA}$ of preamplifier at drive frequency

$s_{uA} = S_{UA}/( I_0 R\Phi_0/2\pi ) = 4\Gamma_A \gamma_{RA} = 4\Gamma_A \beta'_L f_0 Q_A$:  normalized spectral density of short circuit voltage noise $U_{NA}$ of preamplifier at drive frequency

$S_{Ui}$:  spectral density of low-frequency voltage noise of input circuit resistor $R_i$

$s_{ui} = 4\Gamma_i \gamma_{Ri} = 4\Gamma_i \beta'_L f_{0i} / Q_{0i}$:  normalized spectral density of low-frequency voltage noise of input circuit resistor $R_i$

$S_{VA}$:  low-frequency spectral density of short circuit voltage noise of preamplifier

$s_{vA} = S_{VA}/( I_0 R\Phi_0/2\pi ) = = 4\Gamma_A \gamma_{RA} = 4\Gamma_A \beta'_L f_0 Q_A$:  normalized low-frequency spectral density of short circuit voltage noise of preamplifier $S_{VJ}$:  low-frequency cross spectral density between Fourier transforms of low frequency voltage across tank circuit and circulating current in SQUID in the absence of input circuit and noise due to $R_A$ (i. e. $\alpha_i = \alpha_{iT} = 0$, $Q_A \rightarrow \infty$ ).

$s_{vj} = S_{VJ} / I_0 \Phi_0 / 2\pi \gamma_L$:  normalized low-frequency cross spectral density between Fourier transforms of low frequency voltage across tank  circuit and circulating current in SQUID in the absence of input circuit and noise due to $R_A$ (i. e. $\alpha_i = \alpha_{iT} = 0$, $Q_A \rightarrow \infty$ )

$\tilde{s}_{vj}$ : normalized low-frequency cross spectral density between Fourier transforms of low frequency tank circuit voltage and circulating current in SQUID for $\alpha_i = \alpha_{iT} = 0$;

$$\tilde{s}_{vj} = \mathrm{Re}(\tilde{j}_{0\omega})\,\mathrm{Re}(\tilde{v}_{T0\omega}) + \mathrm{Im}(\tilde{j}_{0\omega})\,\mathrm{Im}(\tilde{v}_{T0\omega}) \approx s_{vj0\,vj} + c\Gamma_A / Q_A,$$ where $c$ is a numerical coefficient

$\bar{s}_{vj} = \mathrm{Re}(\tilde{j}_{0\omega})\,\mathrm{Im}(\tilde{v}_{T\omega}) - \mathrm{Im}(\tilde{j}_{0\omega})\,\mathrm{Re}(\tilde{v}_{T\omega})$

$s_{vT0}$:  normalized low-frequency noise spectral density of tank circuit voltage in the absence of input circuit and noise due to $R_A$ (i. e. $\alpha_i = \alpha_{iT} = 0$, $Q_A \rightarrow \infty$ ).

$\tilde{s}_{vT}$ :  normalized noise spectral density of tank circuit voltage in the absence of input circuit ($\alpha_i = \alpha_{iT} = 0$); $\tilde{s}_{vT} = s_{vT0} + a\Gamma_A / Q_A$, where $a$ is a numerical coefficient

$S_{VT}$: spectral density of low frequency voltage noise across the tank circuit in the presence of an input circuit and including the back action of the preamplifier

$s_{vT} = S_{VT}/[I_0 R \Phi_0 / 2\pi]$: spectral density of normalized low frequency voltage noise $S_{VT}$

$T_A$:  effective temperature of preamplifier input resistor $R_A$

$T_i$:  temperature of resistor in input circuit

$T_N$:  noise temperature

$T_{NA} = S_{IA}^{1/2} S_{VA}^{1/2} / 2k_B$ : optimized noise temperature of preamplifier

$T_{N,opt}$:  noise temperature optimized for parameters $\alpha_i$, $f_{0i}$ of tuned input circuit

$T_{N,res}$:  noise temperature for tuned input circuit at resonance

$T_{tank}$: effective temperature of tank circuit resistor, including noise of the preamplifier

$U_A$: voltage across resistor $R_A$

$u_A = U_A / I_0 R$: normalized voltage across resistor $R_A$

$U_{Ci}$: voltage across input circuit capacitor

$u_{Ci} = U_{Ci} / I_0 R$: normalized noise voltage across capacitor $C_i$

$U_{Li}$: voltage across input circuit inductor

$u_{Li} = U_{Li} / I_0 R$: normalized noise voltage across inductor $R_i$

$U_{NA}$: noise voltage added by preamplifier at drive frequency

$u_{NA}$: normalized noise voltage added by preamplifier at drive frequency

$U_{Ni}$: voltage noise source with spectral density $4k_B T_i R_i$

$u_{Ni} = U_i / I_0 R$: normalized noise voltage of resistor $R_i$

$U_{Ri}$: voltage across input circuit resistor

$u_{Ri} = U_{Ri} / I_0 R$: normalized voltage across resistor $R_i$

$V_{NA}$: low frequency noise voltage added by preamplifier

$v_{NA}$: low frequency normalized noise voltage added by preamplifier

$v_{T0}$: normalized low-frequency voltage across tank circuit at optimal bias point in the absence of input circuit and noise due to $R_A$ (i. e. $\alpha_i = \alpha_{iT} = 0$, $Q_A \to \infty$)

$v_{T0\omega}$: Fourier component at frequency $\omega$ of $v_{T0}$

$\tilde{v}_{T0\omega}$: Fourier component at frequency $\omega$ of normalized low frequency voltage across tank circuit for $\alpha_i = \alpha_{iT} = 0$

$V_\varphi$: Modulus of transfer function $dV_T / d\Phi_{ext}$

$v_\varphi = | dv_T / d\varphi_{ext} |$: normalized modulus of transfer function

$\alpha_i = M_i / \sqrt{L_i L}$: coupling coefficient between input circuit and SQUID

$\alpha_{iT} = M_{iT} / \sqrt{L_i L_T}$: coupling coefficient between input circuit and tank circuit

$\gamma_{Li} = \sqrt{L_i / L}$ : inductance scaling parameter between SQUID and input circuit;

throughout the manuscript calculations are for $\gamma_{Li} = 1$;

scaling: $i_i(\gamma_{Li}) = i_i(\gamma_{Li} = 1) / \gamma_{Li}$, $u_i(\gamma_{Li}) = \gamma_{Li} \cdot u_i(\gamma_{Li} = 1)$, $\varphi_i(\gamma_{Li}) = \gamma_{Li} \cdot \varphi_i(\gamma_{Li} = 1)$

$\gamma_{Ri} = R_i / R = \beta'_L f_{0i} / Q_{0i}$ : ratio between input resistor and junction resistor

$\gamma_{RA} = R_A / R = \beta'_L f_0 \gamma_L^2 Q_A$ : ratio between amplifier resistor and junction resistance

$\Gamma_A = 2\pi k_B T_A / I_0 \Phi_0$ : noise parameter related to resistor $R_A$

$\Gamma_i = 2\pi k_B T_i / I_0 \Phi_0$ : noise parameter related to resistor $R_i$

$\Gamma_{N,opt} = 2\pi k_B T_{N,opt} / I_0 \Phi_0$ : normalized noise temperature optimized for parameters of input circuit

$\Gamma_{NA} = 2\pi k_B T_{NA} / I_0 \Phi_0$ : normalized optimal noise temperature of preamplifier

$\Gamma_{N,res} = 2\pi k_B T_{N,res} / I_0 \Phi_0$ : normalized noise temperature for input circuit at resonance

$\Phi_i = L_i I_i + M_i J + M_{iT} I_1$ : flux through input circuit inductor

$\varphi_i = 2\pi \Phi_i / \Phi_0$ : normalized flux through input circuit inductor

$\Phi_T = L_T I_1 + M J + M_{iT} I_i$ : flux through tank circuit inductor

$\varphi_T = 2\pi \Phi_T / \Phi_0$ : normalized flux through tank circuit inductor

$\kappa = R_{opt} [R_A^{-1} + (Q_0^2 R_T)^{-1}]$ : ratio of optimal resistance of preamplifier to tank circuit impedance at resonance

$\xi = 2 Q_{eff,0} (f_d / f_0 - 1)$ : detuning parameter

**Figure Captions**

Fig. 1. Rf SQUID circuit including the SQUID, an input circuit and a readout tank circuit. The resistor $R_A$ in parallel with the tank circuit represents the input resistance of the preamplifier. The arrow between $U_T$ and $V_T$ indicates down-conversion of the voltage across the tank circuit.

Fig. 2. Noise power $\tilde{s}_j$ (in units of $I_0 R \Phi_0 / 2\pi$) of the circulating current $j$ in the rf SQUID together with the noise power $s_{uT}$ (in units of $I_0 \Phi_0 / 2\pi R$) of the voltage $u_T$ across the tank circuit. In (a) the parameters are chosen to optimize the noise temperature, in (b) to optimize the noise energy. The spectra have been averaged 100 times.

Fig. 3. Optimized values of $\Gamma_{N,opt} / f$ as a function of $\alpha$ for $Q_0 = 100$, $Q_A \rightarrow \infty$ (squares) and for $Q_0 = Q_A = 200$, $\Gamma_A = 0.25$ (circles). Other parameters are indicated in the figure. Selected model parameters are listed in Table 1.

Fig. 4. $\Gamma_{N,opt} / f$, optimized for $i_d, f_d$ and $\alpha$, vs. $\beta'_L$. Selected model parameters are listed in Table 2.

Fig. 5. Spectral densities $\tilde{s}_{vT}$, $\tilde{s}_j$ and $\tilde{s}_{vj}$, modulus of transfer function $v_\varphi$ and $\Gamma_{N,opt} / f = \pi \sqrt{s_{vT} s_j - s_{vj}^2} / v_\varphi$ (a) vs. $\Gamma$ for $\Gamma_A = 0$ and (b) vs. $\Gamma_A / Q_A$ for $Q_A = 200$ and $\Gamma = 0$. Other model parameters are $Q_0 = 200$ $\alpha_i = \alpha_{iT} = 0$, $f_0 = 0.1$, $\varphi_{ext} = 0.25$ and $\beta_c = 0$. Selected model parameters are listed in Table 3.

Fig. 6. (a) Spectral densities $s_{uT}$, $\tilde{s}_j$, $\tilde{s}_{vT}$ and $\tilde{s}_{vj}$, and (b) calculated values of $\Gamma_{N,opt}/f$, $\Gamma_{N,res}/f$, $e$ and $\alpha_i^2 Q_i\big|_{res}$ vs. frequency for a readout scheme with bandwidth $f_d/8$. Model parameters are indicated in the graphs.

Fig. 7. (a) Spectral densities $s_{uT}$, $\tilde{s}_j$, $\tilde{s}_{vT}$ and $\tilde{s}_{vj}$, and (b) calculated values of $\Gamma_{N,opt}/f$, $\Gamma_{N,res}/f$, $e$ and $\alpha_i^2 Q_i\big|_{res}$ vs. frequency for a readout scheme with bandwidth $f_d/8$. The quantity $\Gamma_{N,opt}/f$, averaged over the full bandwidth, has been optimized with respect to $i_d$, $f_d$, $\alpha$, $\varphi_{ext}$ and $\beta_L'$. Model parameters are indicated in (a).

Fig. 8. (a) Spectral densities $s_{uT}$, $\tilde{s}_j$, $\tilde{s}_{vT}$ and $\tilde{s}_{vj}$, and (b) calculated values of $\Gamma_{N,opt}/f$, $\Gamma_{N,res}/f$, $e$ and $\alpha_i^2 Q_i\big|_{res}$ vs. frequency for $f_0 = 0.5$ and a bandwidth $f_d/8$. Model parameters are indicated in (a). In (a) the grey curves on the high frequency side have been calculated for a narrow-band readout for comparison.

Fig. 9. Dependence of $\Gamma_{N,opt} / f$ on the tank circuit resonance frequency $f_0$. Open squares and black circles correspond to the case $Q_0 = 100$, $Q_A \to \infty$. Open circles and grey circles correspond to the case $Q_0 = Q_A = 200$, $\Gamma_A = 0.25$. In all cases $f_d$, $i_d$ and $\beta_L'$ have been optimized. For the open squares $\alpha$ has been optimized as well while for the other cases $\alpha = 0.2$. Open circles are calculated in the absence of the preamplifier voltage noise while grey circles are calculated for $\kappa = 0.2$ and $\Gamma_{NA} = 0.05$. Dotted line is function $0.043 / f_0^{0.5}$, dashed line is function $0.037 / f_0 + 0.1 / f_0^{0.5}$. Selected model parameters are listed in Table 4.

Fig. A1. Low frequency equivalent circuit for the rf SQUID coupled to a tuned input circuit.

Fig. A2. Inset shows a comparison between the voltage noise spectral density calculated via Eq. (A2) and via numerical simulations ("num") of the full circuit. Main panel is the signal-to-noise ratio (SNR) at $\Gamma_i = 10^{-5}$ calculated from these data. Model parameters are $\alpha_i = 0.2$, $\alpha_{iT} = 0$, $\alpha = 0.725$, $Q_{0i} = 50$, $Q_A = 10100$, $Q_0 = 101$, $f_0 = 0.1$, $f = 0.1066$, $\varphi_{ext} = 0.25$, $\Gamma = 0.025$, $\Gamma_A = 0$, $\beta'_L = 1.21$, $\beta_c = 0$ and $i_d = 0.369$. Spectra were averaged 100 times. For the two values of $\Gamma_i$ the same sequence of random numbers was used to provide reasonably smooth SNRs.

Fig. A3. A comparison between the signal-to-noise ratio (SNR) at $\Gamma_i = 10^{-5}$ calculated via Eq. (A2) and via numerical simulations of the full circuit for the model parameters indicated in the figures. Other parameters are $\alpha = 0.725$, $f_0 = 0.1$, $f_d = 0.1066$, $\varphi_{ext} = 0.25$, $\Gamma = 0.025$, $\beta'_L = 1.216$, $\beta_c = 0$ and $i_d = 0.369$. For the solid grey curve in (b) the drive frequency has been increased to $f_d = 0.1088$. All spectra were averaged 100 times. For the two values of $\Gamma_i$ required to calculate the SNR the same sequence of random numbers was used to provide reasonably smooth SNRs.

Fig. A4. Spectral densities $\tilde{s}_{vT}$, $\tilde{s}_j$ and $\tilde{s}_{vj}$, modulus of transfer function $v_\varphi$ and $\Gamma_{N,opt} / f = \pi \sqrt{\tilde{s}_{vT} \tilde{s}_j - \tilde{s}_{vj}^2} / v_\varphi$ vs. $\Gamma_A / Q_A$ for $\Gamma_A = 0.25$ and $Q_{eff,0} = 100$, $\Gamma = 0$, $\alpha_i = \alpha_{iT} = 0$. Quality factor $Q_A$ runs between 10100 and 101. Other model parameters are $f_0 = 0.1$, $\varphi_{ext} = 0.25$, $\beta_c = 0$, $f_d = 0.108$, $i_d = 0.419$, $\alpha = 0.763$ and $\beta'_L = 1.216$.

Fig. A5. Spectral densities $\tilde{s}_{vT}$, $\tilde{s}_{j}$ and $\tilde{s}_{vj}$, modulus of transfer function $v_\varphi$ and $\Gamma_{N,opt}/f = \pi\sqrt{\tilde{s}_{vT}\tilde{s}_{j} - \tilde{s}_{vj}^2}/v_\varphi$ (a) vs. $\Gamma = \Gamma_T$ for $\Gamma_A = 0$, (b) vs. $\Gamma_A/Q_A$ with $Q_A = 200$ and $\Gamma = 0$ and (c) vs. $\Gamma = \Gamma_T$ for $\Gamma_A = 1$. Other model parameters are $Q_0 = 200$, $\alpha_i = \alpha_{iT} = 0$, $f_0 = 0.1$, $\varphi_{ext} = 0.25$, $\beta_c = 0$, $f_d = 0.108$, $i_d = 0.419$, $\alpha = 0.763$ and $\beta'_L = 1.216$.

Fig. A6. A comparison between the signal-to-noise ratio (SNR) at (a) $\Gamma_i = 1.5 \times 10^{-5}$ and (b) $\Gamma_i = 2 \times 10^{-5}$, calculated via Eq. (A2) and via numerical simulations of the full circuit for the model parameters indicated in the figures. Other parameters are $\alpha = 0.7625$, $f_0 = 0.1$, $f_d = 0.1081$, $\varphi_{ext} = 0.25$, $\Gamma = 0.025$, $\beta'_L = 1.216$, $\beta_c = 0$ and $i_d = 0.419$. For the solid grey curve in (b) the drive frequency has been increased to 0.1103. All spectra were averaged 100 times. For the two values of $\Gamma_i$ required to calculate the SNR the same sequence of random numbers was used to provide a reasonably smooth SNRs.

Fig. A7 (a) A comparison for the untuned case between the voltage noise spectral density calculated via Eq. (A14) and via numerical simulations (b) The signal-to-noise ratio (SNR) at $\Gamma_i = 5 \times 10^{-4}$ calculated from these data. Model parameters are $\alpha_i = 0.2$, $\alpha_{iT} = 0$, $\alpha = 0.725$, $Q_{0i} = 50$, $Q_A = 10100$, $Q_0 = 101$, $f_0 = 0.1$, $f = 0.1066$, $\varphi_{ext} = 0.25$, $\Gamma = 0.025$, $\beta'_L = 1.216$, $\beta_c = 0$ and $i_d = 0.369$. Spectra were averaged 100 times. The grey curve in (b) is a 10 point floating average over the numeric data.

**Table captions**

Table 1.  Selected parameter values and some resulting quantities for the graphs in Fig. 3. Fixed parameters listed: $\alpha$ and $Q_0$. Other fixed parameters are $Q_{eff,0} = 100$, $f_0 = 0.1$, $\alpha_i = \alpha_{iT} = 0$, $\beta_c = 0$, $\varphi_{ext} = 0.25$, $\Gamma = 0.025$ and $\Gamma_A = 0.25$. Optimized parameters: $i_d$, $\xi = 2Q_{eff,0}(f_d/f_0-1)$, and $\beta'_L$. Resulting quantities: $\tilde{s}_{vT}$, $\tilde{s}_j$, $\tilde{s}_{vj}$, $v_\varphi$, $\alpha_i^2 Q_{0i}$, $\Gamma_{N,res}/f = \pi\sqrt{\tilde{s}_{vT}\tilde{s}_j}/v_\varphi$ and $\Gamma_{N,opt}/f$.

Table 2.  Selected parameter values and some resulting quantities for the graphs in Fig. 4. Fixed parameters listed: $\beta'_L$, $Q_A$ and $\alpha$ (for $\alpha = 0.2$). Other fixed parameters are $Q_{eff,0} = 100$, $f_0 = 0.1$, $\alpha_i = \alpha_{iT} = 0$, $\beta_c = 0$, $\varphi_{ext} = 0.25$, $\Gamma = 0.025$ and $\Gamma_A = 0.25$. Optimized parameters: $\alpha$ (for $\alpha \neq 0.2$), $i_d$, $\xi$, and $\beta'_L$. Resulting quantities: $\tilde{s}_{vT}$, $\tilde{s}_j$, $\tilde{s}_{vj}$, $v_\varphi$, $e, \alpha_i^2 Q_{0i}$, $\Gamma_{N,res}/f = \pi\sqrt{\tilde{s}_{vT}\tilde{s}_j}/v_\varphi$ and $\Gamma_{N,opt}/f$.

Table 3.  Selected parameter values and some resulting quantities for the graphs in Fig. 5. Fixed parameters listed: $\Gamma$, $\Gamma_A$ $\beta'_L$, $Q_A$ and $\alpha$ (for $\alpha = 0.2$). Other fixed parameters are $Q_A = Q_0 = 200$ $\alpha_i = \alpha_{iT} = 0$, $f_0 = 0.1$, $\varphi_{ext} = 0.25$ and $\beta_c = 0$. Optimized parameters: $\alpha$, $\beta'_L$, $i_d$ and $\xi$. Resulting quantities: $\tilde{s}_{vT}$, $\tilde{s}_j$, $\tilde{s}_{vj}$, $v_\varphi$ and $\Gamma_{N,opt}/f$.

Table 4.  Selected parameter values and some resulting quantities for the graphs in Fig. 9. Fixed parameters listed: $f_0$, $Q_A$, $\kappa$, and $\alpha$ (for $\alpha = 0.2$). For $\kappa = 0$ the preamplifier voltage noise has not been included in the calculation of $\Gamma_{N,opt/f}$. For $\kappa = 0$, $\Gamma_{NA} = 0.05$. Other fixed parameters are $\Gamma = 0.025$, $\Gamma_A = 0.25$, $Q_{eff,0} = 100$, $\alpha_i = \alpha_{iT} = 0$, $\varphi_{ext} = 0.25$ and $\beta_c = 0$. Optimized parameters: $\alpha$, $\beta'_L$, $i_d$ and $\xi$. Resulting quantities: $\tilde{s}_{vT}$, $\tilde{s}_j$, $\tilde{s}_{vj}$, $v_\varphi$ and $\Gamma_{N,opt}/f$.

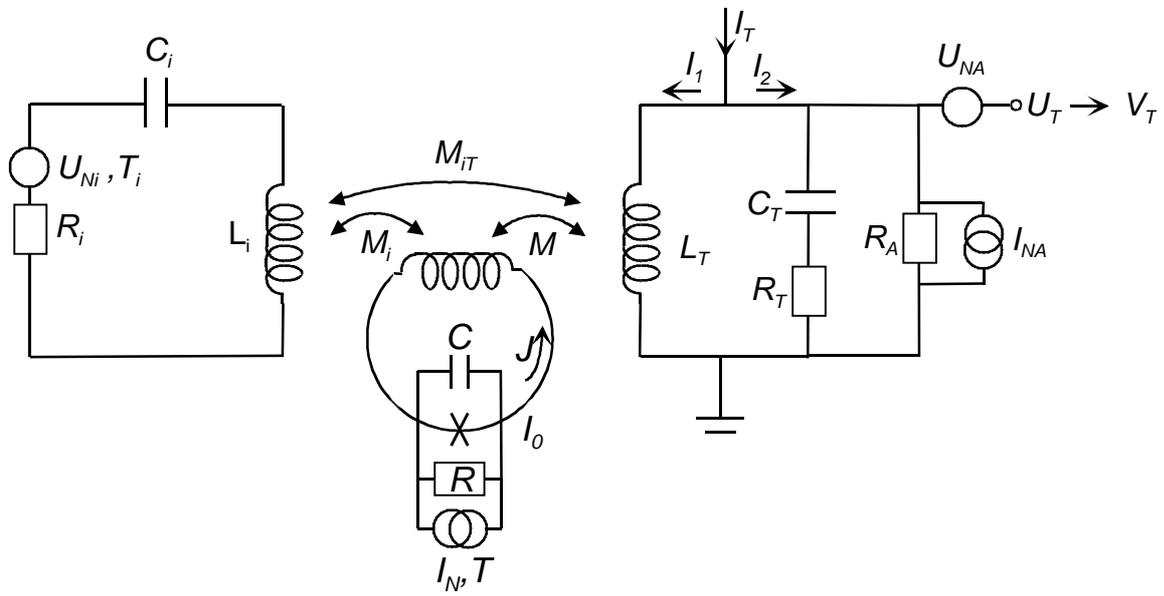

**Figure 1**

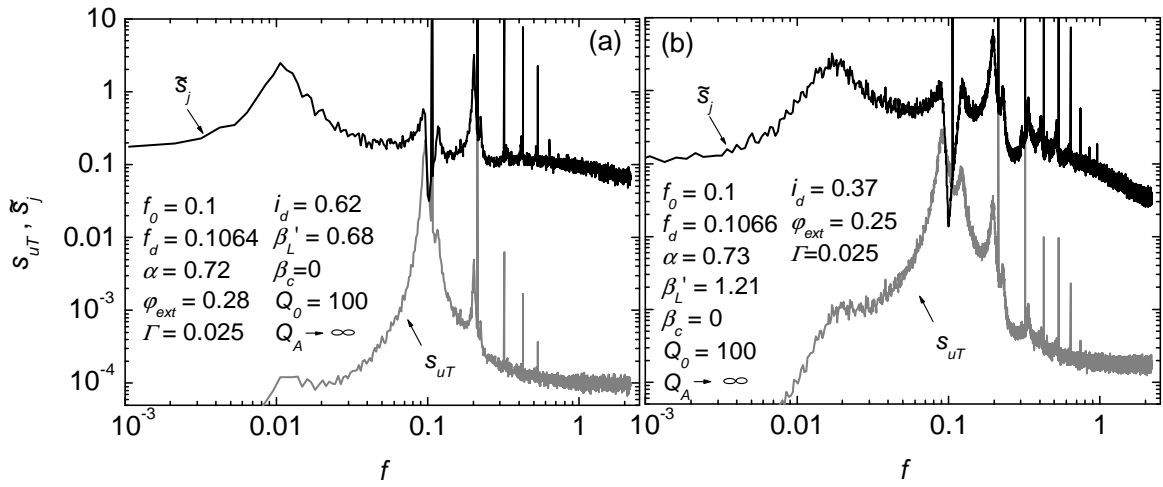

**Figure 2**

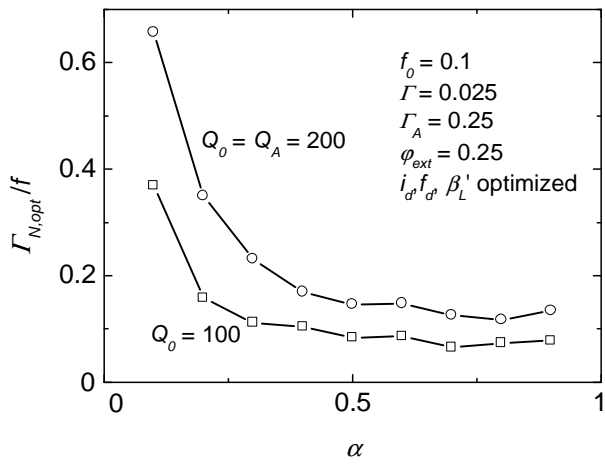

**Figure 3**

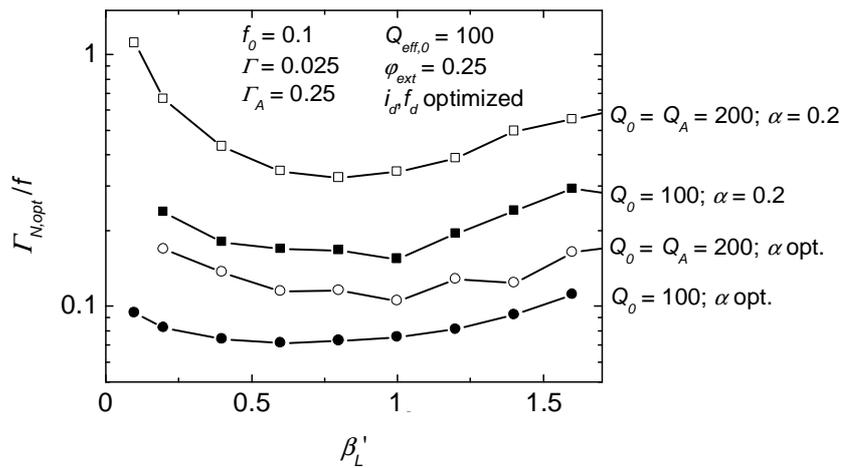

**Figure 4**

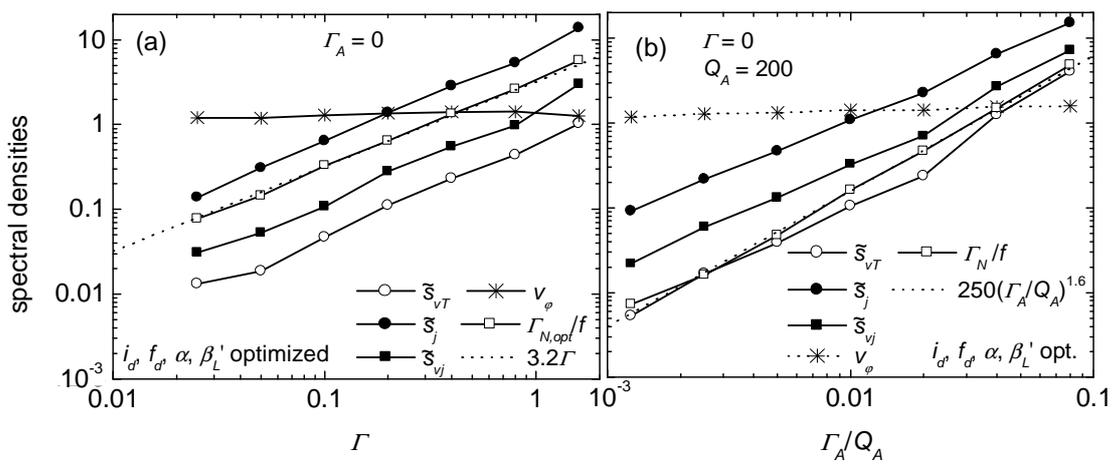

**Figure 5**

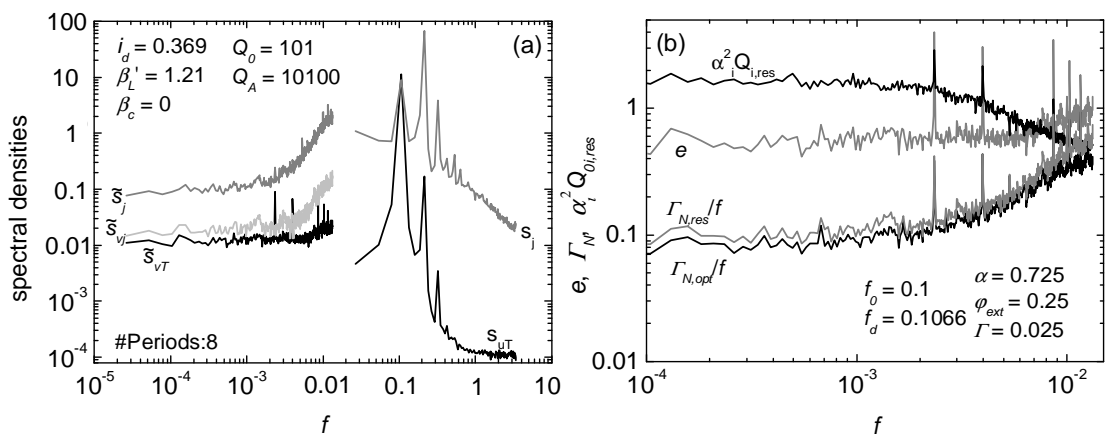

**Figure 6**

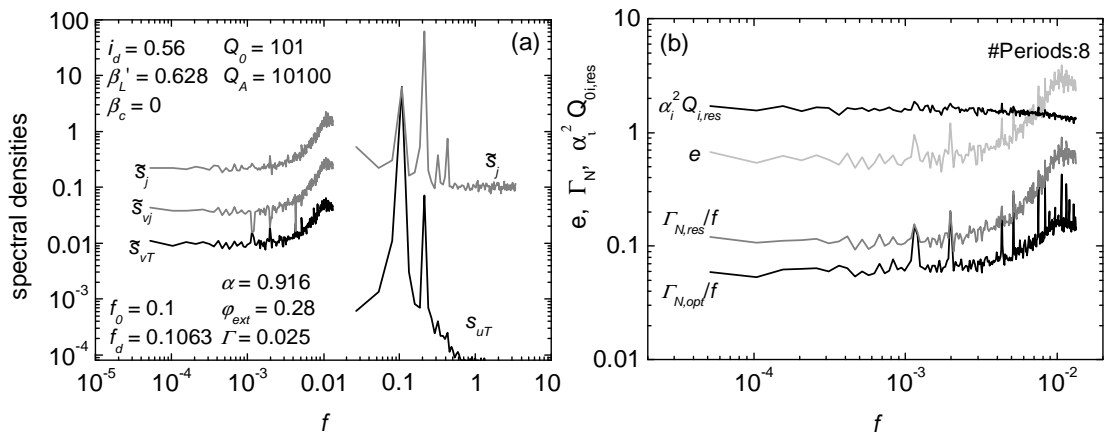

**Figure 7**

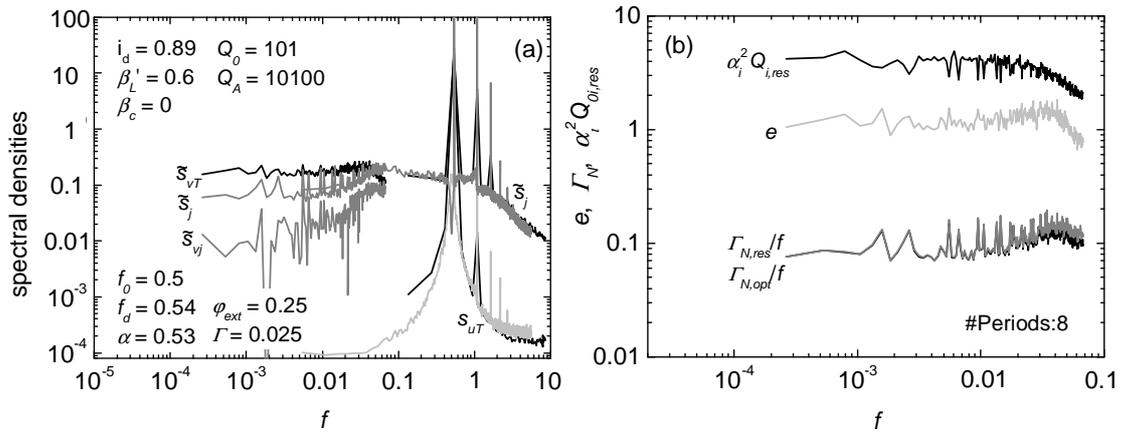

**Figure 8**

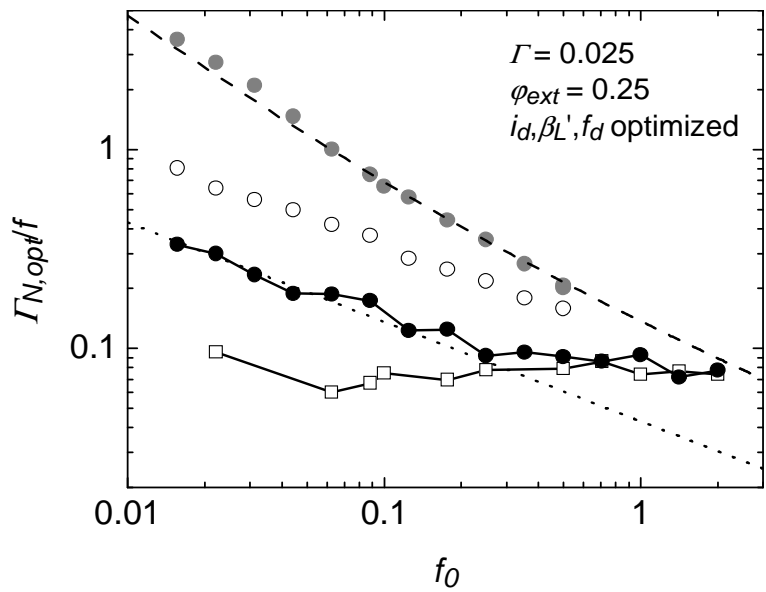

**Figure 9**

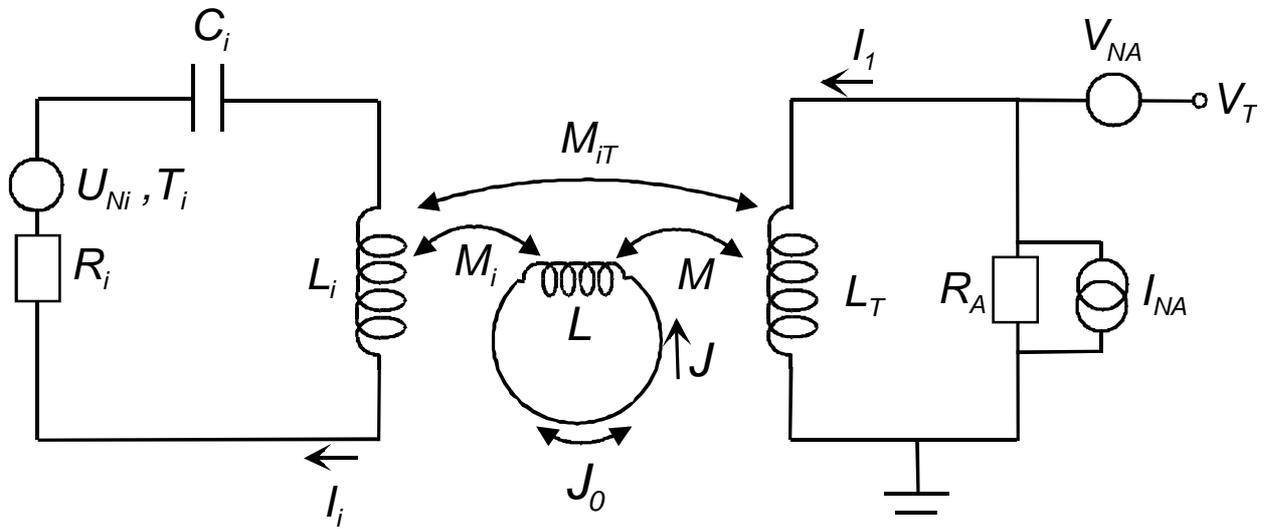

**Figure A1**

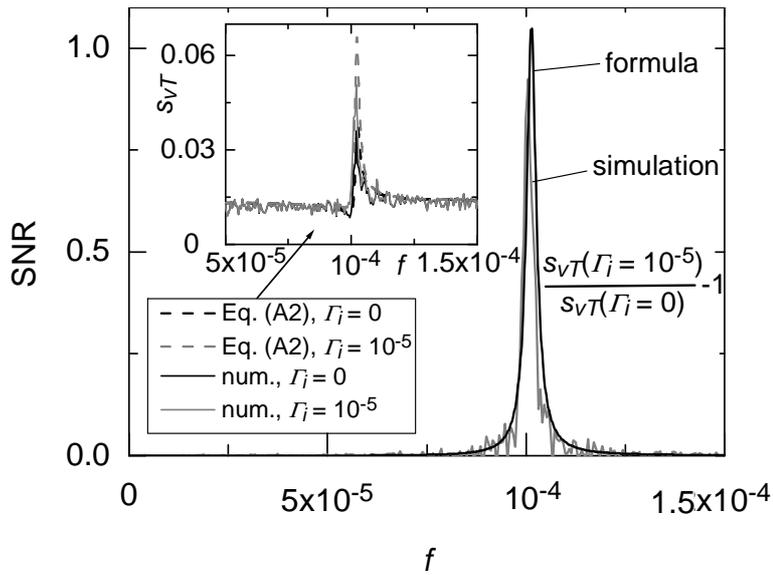

**Figure A2**

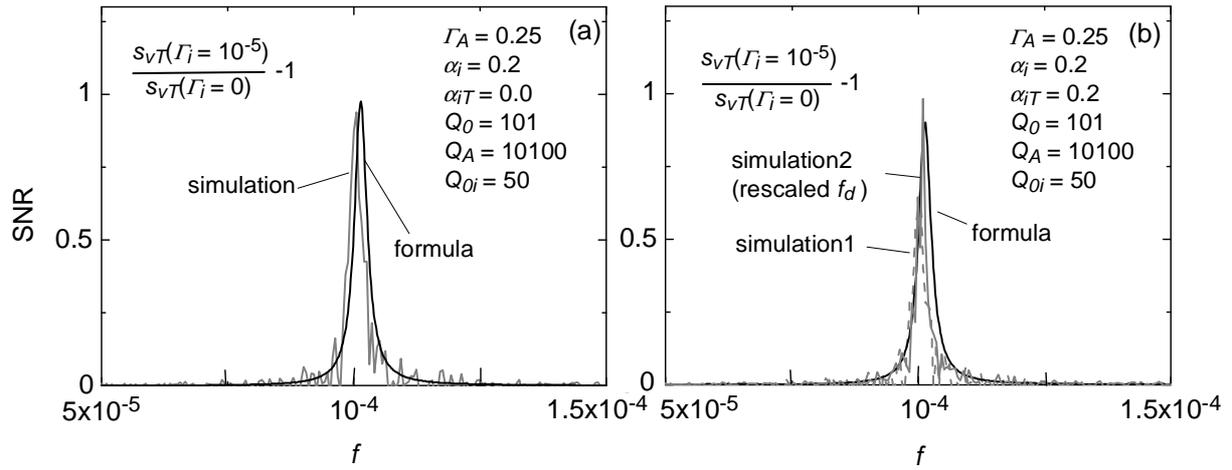

**Figure A3**

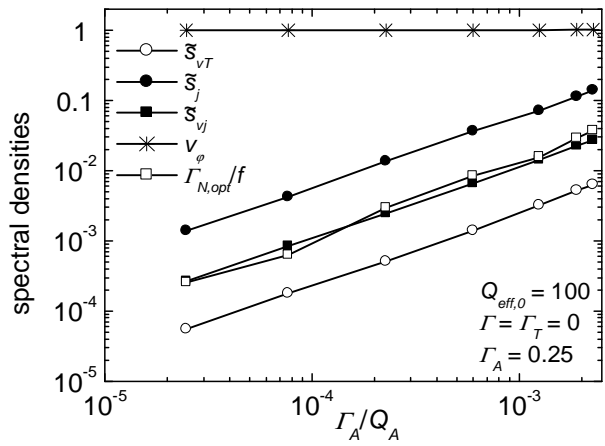

**Figure A4**

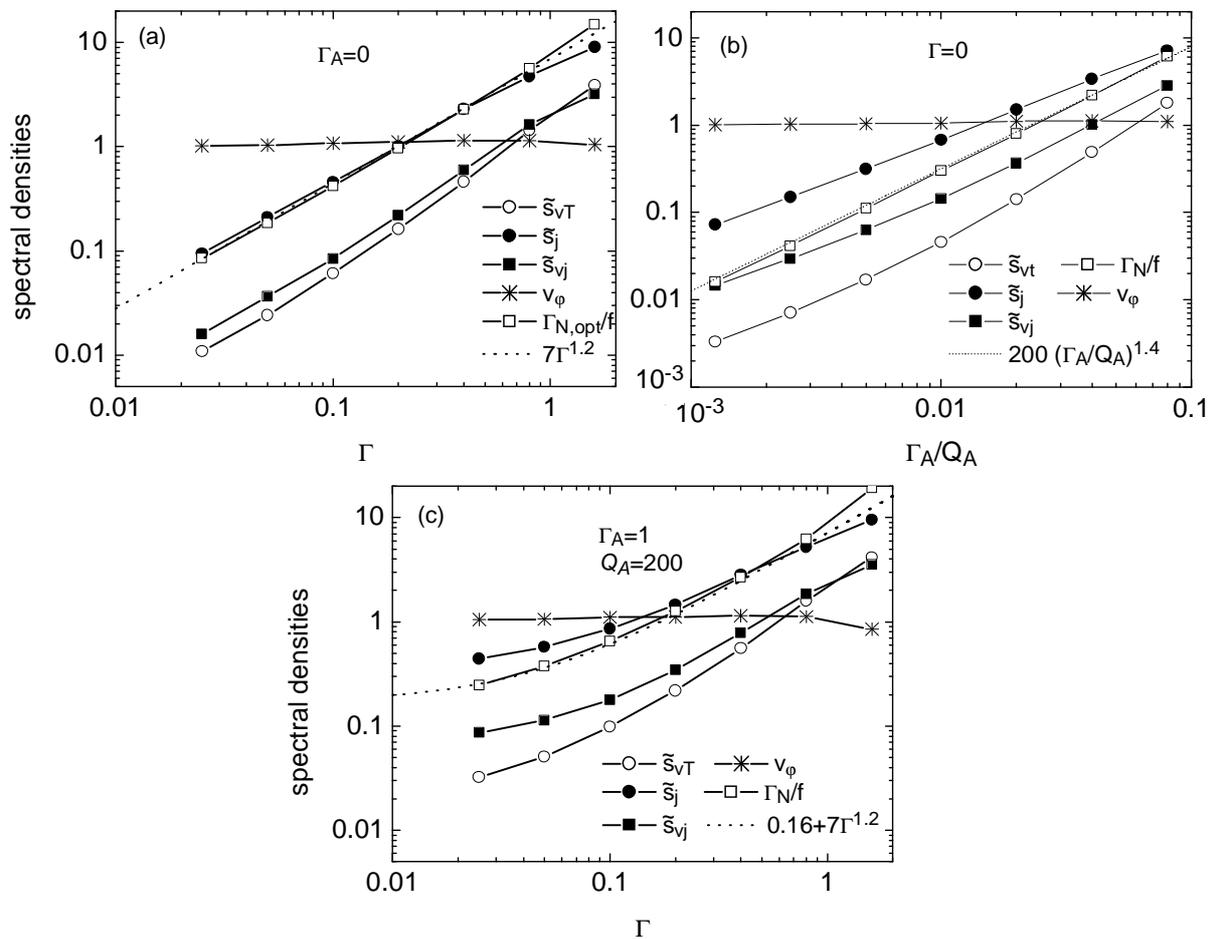

**Figure A5**

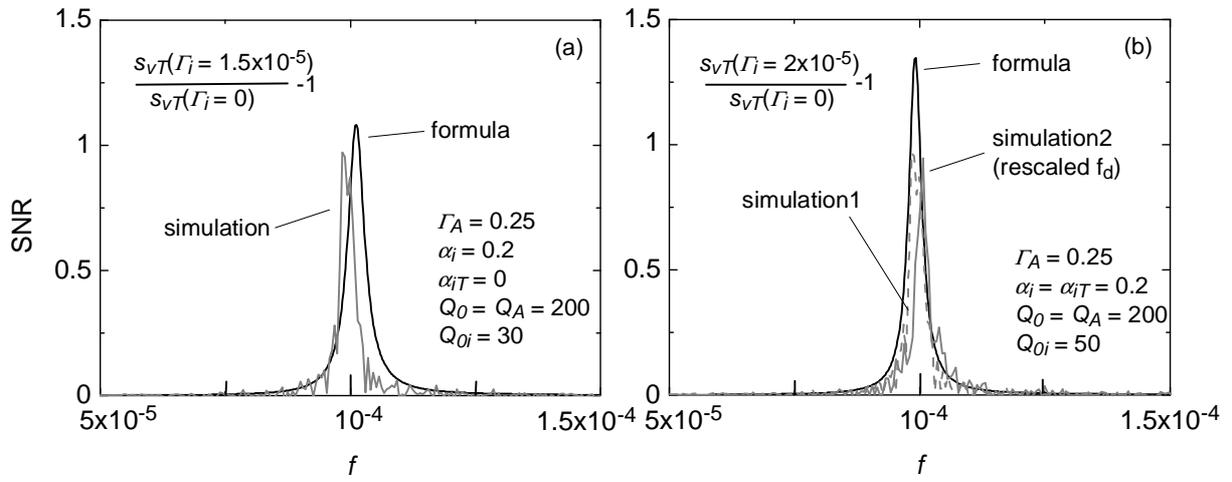

**Figure A6**

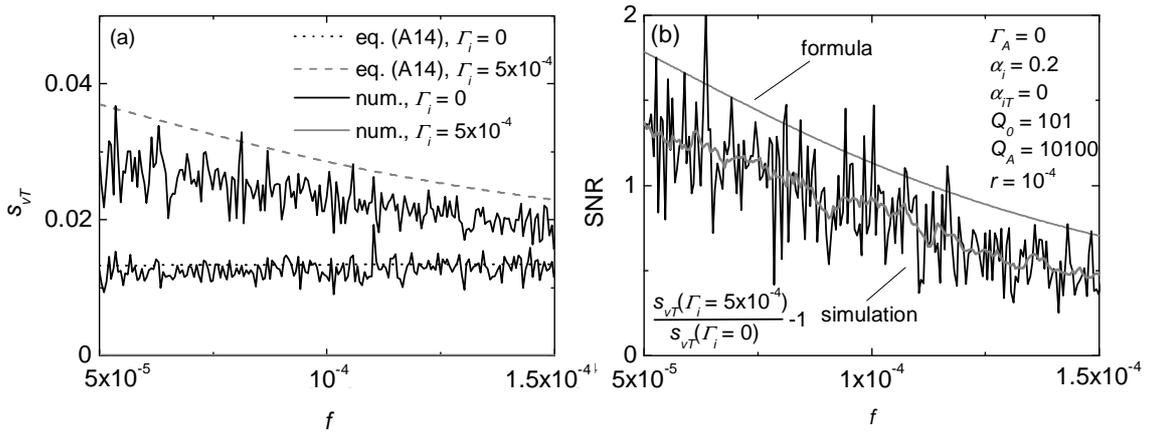

**Figure A7**

| $\alpha$ | $Q_0$ | $i_d$ | $\xi$ | $\beta'_L$ | $\tilde{s}_{vT}$ | $\tilde{s}_j$ | $\tilde{s}_{vj}$ | $v_\varphi$ | $\alpha_i^2 Q_{0i}$ | $\Gamma_{N,res}/f$ | $\Gamma_{N,opt}/f$ |
|---|---|---|---|---|---|---|---|---|---|---|---|
| 0.9 | 100 | 0.72 | 21.8 | 0.58 | 0.0069 | 0.16 | 0.022 | 0.97 | 6.3 | 0.106 | 0.078 |
| 0.7 | 100 | 0.97 | 19.3 | 0.59 | 0.0067 | 0.1 | 0.015 | 1.03 | 6.9 | 0.080 | 0.065 |
| 0.5 | 100 | 1.22 | 21.9 | 0.56 | 0.0057 | 0.08 | 0.005 | 0.80 | 15.6 | 0.085 | 0.083 |
| 0.4 | 100 | 0.80 | 16.9 | 0.72 | 0.013 | 0.08 | 0.006 | 0.95 | 13.4 | 0.106 | 0.10 |
| 0.2 | 100 | 0.67 | 5.9 | 0.81 | 0.059 | 0.16 | 0.022 | 1.88 | 6.7 | 0.16 | 0.16 |
| 0.1 | 100 | 0.81 | 4.6 | 0.82 | 0.12 | 0.13 | 0.019 | 1.05 | 18.4 | 0.37 | 0.37 |
| 0.9 | 200 | 0.65 | 7.45 | 0.31 | 0.033 | 0.84 | 0.15 | 1.36 | 2.9 | 0.38 | 0.13 |
| 0.7 | 200 | 0.55 | 10.4 | 0.53 | 0.025 | 0.44 | 0.091 | 1.32 | 2.2 | 0.25 | 0.13 |
| 0.5 | 200 | 0.6 | 10.4 | 0.71 | 0.036 | 0.31 | 0.082 | 1.45 | 2.1 | 0.23 | 0.15 |
| 0.4 | 200 | 0.63 | 10.4 | 0.8 | 0.05 | 0.25 | 0.077 | 1.5 | 2.3 | 0.23 | 0.17 |
| 0.2 | 200 | 0.63 | 6.6 | 0.92 | 0.14 | 0.45 | 0.13 | 1.98 | 1.9 | 0.4 | 0.35 |
| 0.1 | 200 | 0.75 | 3.7 | 0.87 | 0.49 | 0.63 | 0.36 | 2.01 | 3.2 | 0.86 | 0.66 |

Table 1

| $\beta'_L$ | $Q_A$ | $\alpha$ | $i_d$ | $\xi$ | $\tilde{s}_{vT}$ | $\tilde{s}_j$ | $\tilde{s}_{vj}$ | $v_\varphi$ | $e$ | $\alpha_i^2 Q_{0i}$ | $\Gamma_{N,res}/f$ | $\Gamma_{N,opt}/f$ |
|---|---|---|---|---|---|---|---|---|---|---|---|---|
| 0.1 | $\infty$ | 0.92 | 1.07 | -4.2 | 0.025 | 0.76 | 0.13 | 1.41 | 7.79 | 8.0 | 0.31 | 0.094 |
| 0.4 | $\infty$ | 0.86 | 0.55 | 7.3 | 0.017 | 0.36 | 0.07 | 1.42 | 1.31 | 2.4 | 0.17 | 0.074 |
| 0.6 | $\infty$ | 0.77 | 0.47 | 12.3 | 0.011 | 0.18 | 0.034 | 1.22 | 0.74 | 2.1 | 0.11 | 0.071 |
| 1 | $\infty$ | 0.73 | 0.62 | 19.3 | 0.01 | 0.1 | 0.019 | 1.09 | 0.54 | 1.8 | 0.09 | 0.075 |
| 1.6 | $\infty$ | 0.68 | 0.32 | 18.1 | 0.015 | 0.091 | 0.013 | 0.97 | 0.62 | 1.6 | 0.12 | 0.111 |
| 0.2 | $\infty$ | 0.2 | 1.12 | 2.1 | 0.085 | 0.19 | 0.096 | 1.09 | 22.7 | 19.6 | 19.6 | 0.24 |
| 0.4 | $\infty$ | 0.2 | 0.87 | 4.4 | 0.050 | 0.14 | 0.05 | 1.18 | 5.6 | 7.9 | 7.9 | 0.18 |
| 0.6 | $\infty$ | 0.2 | 0.77 | 4.5 | 0.069 | 0.17 | 0.057 | 1.69 | 2.5 | 4.0 | 4.0 | 0.17 |
| 1.2 | $\infty$ | 0.2 | 0.65 | 8.2 | 0.097 | 0.24 | -0.073 | 2.15 | 1.1 | 1.6 | 1.6 | 0.19 |
| 1.4 | $\infty$ | 0.2 | 0.77 | 9.7 | 0.109 | 0.21 | -0.067 | 1.75 | 1.6 | 1.9 | 1.9 | 0.24 |
| 1.6 | $\infty$ | 0.2 | 0.72 | 9.2 | 0.140 | 0.19 | -0.068 | 1.58 | 2.2 | 2.1 | 2.1 | 0.29 |
| 2 | $\infty$ | 0.2 | 0.47 | 6.5 | 0.37 | 0.12 | -0.081 | 1.91 | 3.2 | 2.8 | 2.8 | 0.33 |
| 0.2 | 200 | 0.8 | 0.48 | 4.25 | 0.067 | 1.45 | 0.30 | 1.56 | 8.7 | 9.52 | 0.63 | 0.17 |
| 0.4 | 200 | 0.75 | 0.67 | 7.75 | 0.04 | 0.69 | 0.15 | 1.47 | 2.9 | 5.45 | 0.36 | 0.14 |
| 0.6 | 200 | 0.72 | 0.55 | 9.88 | 0.028 | 0.41 | 0.097 | 1.39 | 1.6 | 4.11 | 0.25 | 0.11 |
| 1.2 | 200 | 0.85 | 0.48 | 13.9 | 0.019 | 0.26 | 0.055 | 1.07 | 0.86 | 2.9 | 0.21 | 0.13 |
| 1.6 | 200 | 0.72 | 0.41 | 22.9 | 0.018 | 0.14 | 0.012 | 0.95 | 0.78 | 5.1 | 0.17 | 0.16 |
| 2 | 200 | 0.57 | 0.4 | 26.5 | 0.106 | 0.094 | 0.015 | 1.18 | 2.42 | 7.0 | 0.27 | 0.26 |
| 0.2 | 200 | 0.2 | 0.68 | 0.97 | 0.88 | 1.34 | 1.01 | 1.39 | 135 | 18 | 2.38 | 0.66 |
| 0.4 | 200 | 0.2 | 0.52 | 1.91 | 0.64 | 1.09 | 0.79 | 2.05 | 23.7 | 5.8 | 1.28 | 0.43 |
| 0.6 | 200 | 0.2 | 0.62 | 3.91 | 0.29 | 0.58 | 0.35 | 1.91 | 8.4 | 3.9 | 0.67 | 0.34 |
| 1.2 | 200 | 0.2 | 0.32 | 3.78 | 0.28 | 0.66 | 0.23 | 2.94 | 1.7 | 1.2 | 0.46 | 0.39 |
| 1.4 | 200 | 0.2 | 0.58 | 2.69 | 0.44 | 0.5 | -0.18 | 2.78 | 2.6 | 1.5 | 0.53 | 0.49 |

Table 2

| $\Gamma$ | $\Gamma_A$ | $\alpha$ | $\beta'_L$ | $i_d$ | $\xi$ | $\tilde{s}_{vT}$ | $\tilde{s}_j$ | $\tilde{s}_{vj}$ | $v_\varphi$ | $\Gamma_{N,opt}/f$ |
|---|---|---|---|---|---|---|---|---|---|---|
| 0.025 | 0 | 0.7 | 0.6 | 0.39 | 13.3 | 0.013 | 0.14 | 0.031 | 1.2 | 0.078 |
| 0.1 | 0 | 0.65 | 0.53 | 0.69 | 11.2 | 0.047 | 0.64 | 0.11 | 1.29 | 0.33 |
| 0.4 | 0 | 0.68 | 0.52 | 0.58 | 9.4 | 0.23 | 2.85 | 0.55 | 1.41 | 1.33 |
| 1.6 | 0 | 0.68 | 0.26 | 0.88 | 6.5 | 1.04 | 13.7 | 2.98 | 1.27 | 5.73 |
| 0 | 0.0025 | 0.66 | 1.05 | 0.6 | 15.2 | 0.017 | 0.22 | 0.061 | 1.28 | 0.016 |
| 0 | 0.01 | 0.56 | 1.15 | 0.45 | 11.2 | 0.11 | 1.08 | 0.33 | 1.42 | 0.16 |
| 0 | 0.04 | 0.59 | 1.06 | 0.45 | 10.2 | 1.25 | 6.51 | 2.7 | 1.55 | 1.49 |
| 0 | 0.08 | 0.6 | 1.06 | 0.4 | 9.2 | 4.06 | 15.2 | 7.3 | 1.59 | 4.85 |

Table 3.

| $f_0$ | $Q_A$ | $\kappa$ | $\alpha$ | $\beta'_L$ | $i_d$ | $\xi$ | $\tilde{s}_{vT}$ | $\tilde{s}_j$ | $\tilde{s}_{vj}$ | $v_\varphi$ | $\Gamma_{N,opt}/f$ |
|---|---|---|---|---|---|---|---|---|---|---|---|
| 0.0884 | $\infty$ | 0 | 0.78 | 0.83 | 0.91 | 27.0 | 0.0045 | 0.086 | 0.008 | 0.84 | 0.07 |
| 0.0625 | $\infty$ | 0 | 0.73 | 0.63 | 0.94 | 21.0 | 0.0023 | 0.096 | 0.0084 | 0.64 | 0.06 |
| 0.0221 | $\infty$ | 0 | 0.91 | 0.69 | 0.72 | 17.0 | 0.0008 | 0.34 | 0.014 | 0.25 | 0.10 |
| 0.5 | $\infty$ | 0 | 0.2 | 0.54 | 0.74 | 3.99 | 0.65 | 0.089 | 0.0006 | 8.31 | 0.09 |
| 0.25 | $\infty$ | 0 | 0.2 | 0.60 | 0.62 | 4.25 | 0.18 | 0.098 | 0.029 | 4.46 | 0.09 |
| 0.125 | $\infty$ | 0 | 0.2 | 0.49 | 0.64 | 3.50 | 0.086 | 0.12 | 0.060 | 2.11 | 0.12 |
| 0.0625 | $\infty$ | 0 | 0.2 | 0.50 | 0.88 | 4.96 | 0.026 | 0.18 | 0.045 | 0.84 | 0.19 |
| 0.03125 | $\infty$ | 0 | 0.2 | 0.48 | 0.83 | 4.16 | 0.013 | 0.28 | 0.049 | 0.47 | 0.23 |
| 0.01563 | $\infty$ | 0 | 0.2 | 0.64 | 0.51 | 4.03 | 0.0078 | 0.59 | 0.060 | 0.30 | 0.33 |
| 0.5 | 200 | 0 | 0.2 | 0.78 | 0.36 | 2.47 | 2.86 | 0.20 | 0.20 | 13.0 | 0.16 |
| 0.25 | 200 | 0 | 0.2 | 0.80 | 0.34 | 3.26 | 1.00 | 0.28 | 0.28 | 5.94 | 0.22 |
| 0.125 | 200 | 0 | 0.2 | 0.77 | 0.52 | 4.61 | 0.27 | 0.43 | 0.43 | 2.62 | 0.28 |
| 0.0625 | 200 | 0 | 0.2 | 0.78 | 0.56 | 4.89 | 0.12 | 0.84 | 0.84 | 1.26 | 0.42 |
| 0.03125 | 200 | 0 | 0.2 | 0.63 | 0.73 | 5.15 | 0.053 | 1.18 | 1.18 | 0.52 | 0.56 |
| 0.01563 | 200 | 0 | 0.2 | 0.77 | 0.69 | 5.72 | 0.0241 | 2.31 | 2.31 | 0.28 | 0.81 |
| 0.5 | 200 | 0.2 | 0.2 | 1.04 | 0.31 | 2.58 | 5.31 | 0.18 | 0.24 | 14.3 | 0.21 |
| 0.25 | 200 | 0.2 | 0.2 | 0.90 | 0.37 | 4.00 | 1.87 | 0.27 | 0.28 | 5.85 | 0.35 |
| 0.125 | 200 | 0.2 | 0.2 | 0.95 | 0.26 | 2.81 | 0.84 | 0.65 | 0.34 | 3.57 | 0.58 |
| 0.0625 | 200 | 0.2 | 0.2 | 0.99 | 0.50 | 4.52 | 0.41 | 0.78 | 0.33 | 1.43 | 1.00 |
| 0.03125 | 200 | 0.2 | 0.2 | 1.17 | 0.34 | 4.13 | 0.23 | 1.79 | 0.29 | 0.85 | 2.10 |
| 0.01563 | 200 | 0.2 | 0.2 | 0.57 | 0.40 | 3.07 | 0.074 | 1.24 | 0.15 | 0.24 | 3.46 |

Table 4